\documentclass[fleqn,usenatbib,useAMS]{mnras}

\usepackage{graphicx}	
\usepackage{amsmath}	
\usepackage{amssymb}	
\usepackage[flushleft]{threeparttable}
\usepackage{booktabs}
\usepackage{bm}         
\usepackage[dvipsnames]{xcolor}
\hypersetup{breaklinks=true,colorlinks=true,linkcolor=NavyBlue!50!Emerald,citecolor=NavyBlue!50!Emerald,filecolor=blue,urlcolor=Gray} 
\usepackage{ulem}

\newcommand{\kms}{\,km\,s$^{-1}$} 

\usepackage[T1]{fontenc}
\usepackage{ae,aecompl}

\usepackage[british]{babel}
\usepackage{newtxtext}
\usepackage[frenchmath,varg]{newtxmath}

\def\kms{\mbox{${\rm km}\:{\rm s}^{-1}\:$}}

\def\lesssim{\mathrel{\hbox{\rlap{\hbox{\lower4pt\hbox{$\sim$}}}\hbox{$<$}}}}
\let\la=\lesssim
\def\gtrsim{\mathrel{\hbox{\rlap{\hbox{\lower4pt\hbox{$\sim$}}}\hbox{$>$}}}}
\let\ga=\gtrsim

\def\farcs{\hbox{$.\!\!^{\prime\prime}$}}

\def\lx{$L_\mathrm{X}$}

\def\ledd{$L_\mathrm{Edd}$}

\title[AGN accretion states]{X-ray binary accretion states in Active Galactic Nuclei? \\ Sensing the accretion disc of supermassive black holes with mid-infrared nebular lines}

\author[Fern\'andez-Ontiveros \& Mu\~noz-Darias]{
Juan A. Fern\'andez-Ontiveros$^{1,}$\thanks{E-mail: \textsf{\href{mailto:j.a.fernandez.ontiveros@gmail.com}{j.a.fernandez.ontiveros@gmail.com}}} \&
Teo Mu\~noz-Darias$^{2,3}$
\vspace{0.2cm}\\
$^1$Istituto di Astrofisica e Planetologia Spaziali (INAF--IAPS), Via Fosso del Cavaliere 100, I--00133 Roma, Italy
\vspace{0.03cm}\\
$^2$Instituto de Astrof\'isica de Canarias (IAC), C/V\'ia L\'actea s/n, E--38205 La Laguna, Tenerife, Spain\\
$^3$Universidad de La Laguna (ULL), Dpto. Astrof\'isica, Avd. Astrof\'isico Fco. S\'anchez s/n, E--38206 La Laguna, Tenerife, Spain
}

\date{Accepted 2021 April 15. Received 2021 April 15; in original form 2021 January 14}

\pubyear{2021}

\begin{document}
\label{firstpage}
\pagerange{\pageref{firstpage}--\pageref{lastpage}}
\maketitle

\begin{abstract}
Accretion states, which are universally observed in stellar-mass black holes in X-ray binaries, might be expected in active galactic nuclei (AGN). This is the case at low luminosities, when the jet-corona coupling dominates the energy output in both populations. Previous attempts to extend this framework to a wider AGN population have been extremely challenging due to heavy hydrogen absorption of the accretion disc continuum and starlight contamination from the host galaxies. We present the luminosity-excitation diagram (LED), based on the [\ion{O}{iv}]$_{\rm 25.9\mu m}$ and [\ion{Ne}{ii}]$_{\rm 12.8\mu m}$ mid-infrared nebular line fluxes. This tool enables to probe the accretion disc contribution to the ionising continuum. When applied to a sample of 167 nearby AGN, the LED recovers the characteristic q-shaped morphology outlined by individual X-ray binaries during a typical accretion episode, allowing us to tentatively identify the main accretion states. The \textit{soft state} would include broad-line Seyferts and about half of the Seyfert 2 population, showing highly excited gas and radio-quiet cores consistent with disc-dominated nuclei, in agreement with previous studies. The \textit{hard state} mostly includes low-luminosity AGN ($\lesssim 10^{-3}\, \rm{L_{Edd}}$) characterised by low-excitation radio-loud nuclei and a negligible disc contribution. The remaining half of Seyfert 2 nuclei and the bright LINERs show low excitation at high accretion luminosities, and could be identified with the bright hard and intermediate states. Their hosts show ongoing star formation in the central kiloparsecs. We discuss the above scenario, its potential links with the galaxy evolution picture and the possible presence of accretion state transitions in AGN, as suggested by the growing population of changing-look quasars.
\end{abstract}




\section{Introduction}\label{intro}

The presence of different accretion states (e.g. \citealt{Miyamoto1992,Miyamoto1993}) is well established in accreting stellar-mass black holes (BHs). These objects are primarily found in low-mass X-ray binaries, where they are fed via an accretion disc \citep{Shakura1973} by material that is pulled-out from their nearby stellar companions. The vast majority of these BH X-ray binaries (BHXBs) spend most part of their lives in a low luminosity phase (quiescence). This is a direct consequence of the mass transfer rate supplied by the companion being too low to sustain the disc in an active, hot phase in the long run. However, since the outer disc progressively builds up, BHXBs eventually show sporadic outbursts (\citealt{Lasota2001} for a review on the disc instability model), when matter is accreted at much higher rates and the X-ray luminosity (\lx; produced in the inner parts of the accretion flow) can increase over a million fold, even approaching the Eddington limit (\ledd). During these events, lasting months to years, galactic BHXBs become among the brightest sources of the X-ray sky, allowing us to perform sensitive and detailed follow-up studies of a single object (i.e. suffering from the very same systematics) over a broad range of accretion luminosities. This has revealed the universal presence of two main X-ray states, \textit{hard} and \textit{soft}, which are entirely defined based on observational properties, but that most likely encode fundamental differences on the physical processes at work and the accretion geometry. Thus, these X-ray states are considered to be the smoking guns of truly different accretion states (see \citealt{Remillard2006, Done2007, Belloni2011, Fender2012, Fender2016} for reviews). 

\begin{figure}
	\includegraphics[width=\columnwidth]{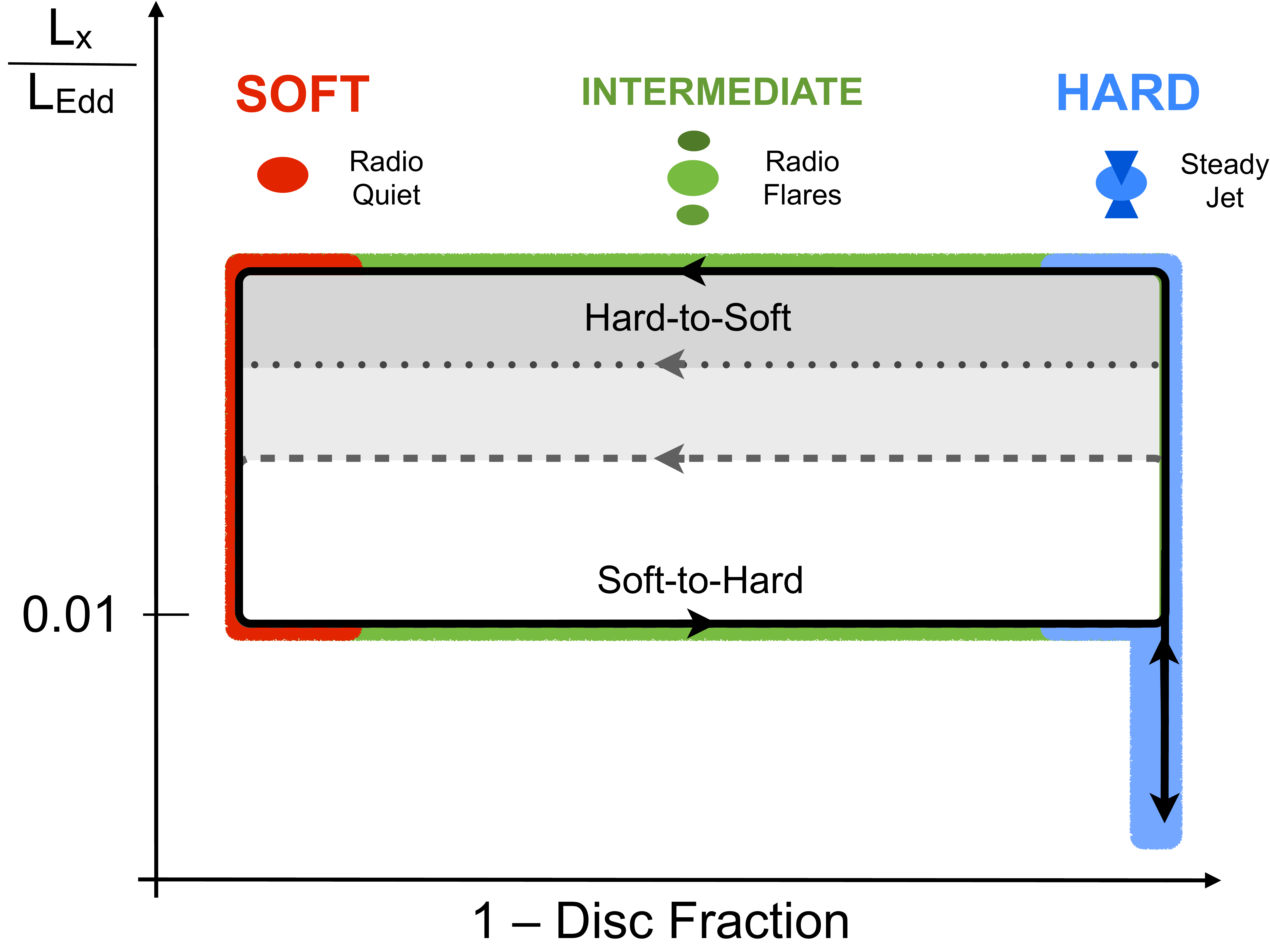}
    \caption{The sketch outlines the (X-ray) evolution of a BHXB during a typical outburst, when the characteristic q-shaped pattern is described (counterclockwise). While the hard-to-soft (high luminosity) transition might happen at different luminosities (even for different outbursts of the same source) the soft-to-hard (low luminosity) transition is observed to occur at roughly the same brightness ($\sim 10^{-2}$\,\ledd). The accretion disc contribution to the X-ray luminosity (disc fraction) can be inferred from spectral fits or, alternatively, traced by observational proxies such as the X-ray colour and the aperiodic variability. The radio (jet) properties of the source at each stage are indicated in the top (see text for details).} \label{fig_BHXBs}
\end{figure}

The above-mentioned states are naturally revealed in the hardness-intensity diagram \citep{Homan2001} or the analogue hardness-luminosity diagram\footnote{A representation of the system luminosity against the X-ray spectral hardness, defined as the ratio of the count rates between two energy bands, typically $0.5-2\, \rm{keV}$ and $2-10\, \rm{keV}$.} (HLD; e.g. \citealt{Dunn2010}). An additional and more physically meaningful version (but model dependent) of the X-ray state picture is provided by the disc fraction luminosity diagram (\citealt{Koerding2006}, see also e.g. \citealt{Kalemci2004}), which was first applied to a large sample of BHXBs by \citet{Dunn2010}. Active BHXBs (\lx $\gtrsim 10^{-4}$\,\ledd) display q-shaped tracks in all these diagrams. At low luminosity ($10^{-4}$\,\ledd\,$\lesssim$ \lx $\lesssim 10^{-2}$\,\ledd) the spectrum is always hard and systems display approximately vertical tracks. This hard state can extend up to a several times $10^{-1}$\,\ledd. Once this high \lx\ regime is reached, a transition to the soft state might occur at relatively constant \lx. When this happens, the spectrum remains soft for a relatively long period of time as \lx\ gradually drops down to $ \sim 10^{-2}$\,\ledd. This is followed by a new transition towards the initial, low-luminosity hard state, from which the system might return to quiescence (see e.g. \citealt{Plotkin2013} for the spectral evolution throughout low \lx/\ledd\ ratios). A sketch describing this evolution is presented in Fig.\,\ref{fig_BHXBs}.

The hard/soft spectral dichotomy is caused by pronounced changes in the balance between two spectral components. While the thermal accretion disc contribution can account for up to 99\% of the total X-ray flux during the soft state (\citealt{Dunn2010}; with steady, long-lasting soft states typically reaching $> 95$\%, \citealt{Munoz-Darias2013b}), this component has a very low temperature (e.g. \citealt{Plant2015}) during the hard state, providing an almost negligible contribution to \lx. The energy spectrum is then dominated by a hard component, that can be fitted by a power law. This likely arises from inverse-Compton processes in an extended hot, optically thin flow (a.k.a. corona; e.g. \citealt{Gilfanov2010}). A significant contribution from Synchrotron (jet) emission and/or a common physical origin for the jet and the Componization components are also plausible scenarios supported by theoretical work and data modelling (e.g. \citealt{1979ApJ...232...34B}, \citealt{Markoff2001}, \citealt{2005ApJ...635.1203M}). Besides systems/outbursts that only sample hard states, the above hard-soft-hard pattern (first described in \citealt{Miyamoto1995}) is always displayed in the same direction (anti-clockwise), with the hard-to-soft transition always occurring at higher \lx. This pattern is hysteresis and it is also observed when looking at the evolution of non-spectral parameters, such as the fast variability \citep{Munoz-Darias2011}. The luminosity of the hard-to-soft transition can vary from system to system, and also between different outbursts of the same BHXB (e.g. \citealt{Dunn2010}). However, the soft-to-hard transitions tend to occur at roughly constant luminosity \citep{Maccarone2003}. Consequently, when plotting (for instance) the disc fraction luminosity diagram for several outburst of several BHXBs, a continuum of data points above $0.01$\,\ledd \ is obtained, covering the full range of disc fractions (i.e. from power law to almost entirely disc dominated; see e.g. fig. 6 in \citealt{Dunn2010}). 
In addition to the widely observed correlation between X-ray spectral and timing properties (including very specific timing features; see e.g. \citealt{Motta2016b}), the strongest piece of evidence connecting the observed X-ray phenomenology with actual accretion states is possibly their coupling with outflows, and in particular with (radio) jets. The hard state is ubiquitously associated with the presence of a compact jet, which is not observed (at least not steadily) in the soft state \citep{Fender2004}. In addition, powerful, discrete ejections are observed during the hard-to-soft transition (see also \citealt{Mirabel1999}; see \citealt{Bright2020} for a recent example). Nevertheless, the most compelling manifestation of the accretion/jet coupling is arguably the universal X-ray/radio correlation found during the hard state (\citealt{Corbel2002,Gallo2003}), which can be extended to radio-loud active galactic nuclei (AGN) when appropriate mass-scaling relations are applied \citep{Merloni2003,Falcke2004}. 

Low mass X-ray binaries with neutron star accretors also show a rather similar accretion/ejection coupling (e.g. \citealt{Miller-Jones2010}), accretion states (e.g. \citealt{vanderKlis2006}) and q-shaped, hysteresis loops in both transient and persistent sources \citep{Munoz-Darias2014}. In addition, a qualitatively similar behaviour has also been observed in accreting white dwarfs (e.g. \citealt{Koerding2008}).

\subsection{The quest for AGN accretion states}
The success of the accretion framework in BHXBs prompted the extension of this scheme across the mass range, motivated by the analogies between accreting stellar-mass and supermassive BHs, such as the presence of accretion discs and radio jets \citep[e.g.][]{Koerding2006}. However, the complex phenomenology observed in AGN turned the generalisation of this accretion scheme into a challenging task. 

Seyfert galaxies have been historically divided into two main classes according to the presence (type 1) or absence (type 2) of broad permitted emission-line transitions in their optical spectra \citep{1971Afz.....7..389K}. Early hypothesis to solve this type 1/type 2 dichotomy were based on \textit{changing look} Seyfert nuclei such as NGC\,4151 \citep{1984MNRAS.211P..33P}, where the appearance and disappearance of the broad lines and the blue featureless continuum were explained as on/off switches of the nuclear source. This idea was largely abandoned when orientation-dependent obscuration effects were found, which led to the Unified Model \citep{1993ARA&A..31..473A}. This is strongly supported by the detection of broad emission lines in a fraction of the type 2 population when looking at the polarised and/or the infrared (IR) spectrum \citep{1985ApJ...297..621A,2003ApJ...583..145T,2005AJ....129.1212Z,2016MNRAS.461.1387R}. However, state transitions resurfaced again over the last years, as the number of changing look AGN was vastly increased by intensive monitoring programs \citep[e.g.][]{2016MNRAS.457..389M}. These fast transitions in the observational properties of the nuclear emission might result from significant physical changes in the surroundings of the central engine, opening a new window for a more accretion-oriented AGN classification scheme that goes beyond the Unified Model. 

The search for AGN accretion states has been traditionally hampered by the intrinsically larger timescales associated with supermassive BHs and the restricted access to the continuum emission arising in the accretion disc (see Section\,\ref{method}). Thus, it does not come as a surprise that the resemblance between stellar-mass and supermassive BHs becomes significantly closer when the disc has a lesser contribution, that is, during the hard state. This is the case of low-luminosity AGN ($L_{\rm bol}/L_{\rm Edd} \lesssim 10^{-3}$; \citealt{1996ApJ...462..183H,1999ApJ...516..672H,Ho2008}), where the disappearance of the blue bump can be associated with the dimming of the accretion disc expected at low accretion rates \citep[e.g.][]{Narayan1994}. The majority of these low-luminosity AGN are radio-loud sources found in elliptical and early-type spiral galaxies \citep{Ho2008}, and most of them fall into the LINER spectral class (Low-Ionisation Nuclear Emission-line Region; \citealt{1980A&A....87..152H}), i.e. their optical spectra show strong emission lines from low-ionisation species. 

At the opposite end, quasars are the most luminous nuclei, able to outshine their host galaxies --\,mostly ellipticals\,-- by orders of magnitude, and are rare in the nearby Universe (3C\,273 is one of the closest at $\sim 730\, \rm{Mpc}$). This allowed previous studies to associate luminous quasars with the soft and intermediate states. Most quasars are radio-quiet nuclei, showing prominent blue bump emission with feeble radio activity, resembling the classical soft state seen in BHXBs (see e.g. \citealt{2020arXiv200407258A}). Radio-loud quasars ($\sim 10\%$ of the quasar population) and bright radio galaxies have been proposed as (hard-to-soft) transition nuclei \citep{Koerding2006}, since they show significant blue bumps in their optical/UV continuum (\citealt{1994ApJS...95....1E}) and/or strong high-excitation lines (\citealt{2019ApJ...870...53B}), suggesting the presence of an accretion disc.

Strictly speaking, there are no significant differences between the brightest Seyfert nuclei and the faintest radio-quiet quasars \citep{1976QJRAS..17..227W,1977ARA&A..15...69W}, which are separated in two classes because of (mainly) historical reasons (see \citealt{1987PASP...99..309L} for a review). However, the situation becomes less clear at lower luminosities ($\sim 10^{43}$--$10^{45}\, \rm{erg\,s^{-1}}$), where the brightness of the host galaxies (typically nearby spirals; \citealt{1943ApJ....97...28S}) rivals that of the Seyfert nuclei. In this scenario, the combined emission can have significant contributions (or be even dominated) by starlight (optical to near-IR) and dust emission from star formation regions (far-IR), even though other spectral ranges can still unveil the AGN contribution (e.g. X-rays in not heavily obscured sources or compact radio cores). As a result, the accretion disc continuum cannot be easily de-blended from the starlight, and it must be probed through indirect methods such as reverberation mapping \citep{1993PASP..105..247P,2018ApJ...857...53C}. 

\subsubsection{Early attempts to build AGN accretion-oriented diagrams}
A pioneering attempt to build an HLD diagram for AGN was presented by \citet{Koerding2006} including quasars from SDSS and a local sample of LINERs. The search for accretion states is addressed using a large statistical sample, based on the premise that the variety of AGN types occupy the main loci of the HLD, yielding results that are broadly consistent with the BHXBs accretion scheme. In particular, they found that bright quasar radio emission is generally associated with lower optical to X-ray flux ratios, which is likely caused by a lower disc contribution. One of the limitations of this study is the lack of BH mass determinations (and thus Eddington-scaled luminosities) for such a large sample of objects, which is expected to introduce a $\sim 2\, \rm{dex}$ scatter. Compatible results were obtained by \citet{2017A&A...603A.127S} using UV and X-ray observations (\textit{XMM-Newton}), while the purely X-ray-based approach attempted by \citet{2018MNRAS.481.3563P} returned more ambiguous results due to lower statistics. A common caveat in the above studies is the difficulty to measure the disc contribution in intermediate-to-low luminosity AGN, i.e. below $\lesssim 10^{43}\, \rm{erg\,s^{-1}}$, since this is heavily absorbed due to the large hydrogen opacity in the Lyman continuum (in addition to contamination from the stellar population in the host galaxy). 

Probing the accretion disc contribution in AGN is the first milestone to test the BHXBs accretion scheme in supermassive BHs. However, direct measurements are only possible for non-obscured and luminous nuclei, where the host galaxy contamination is negligible. 

\begin{figure*}
	\includegraphics[width=\textwidth]{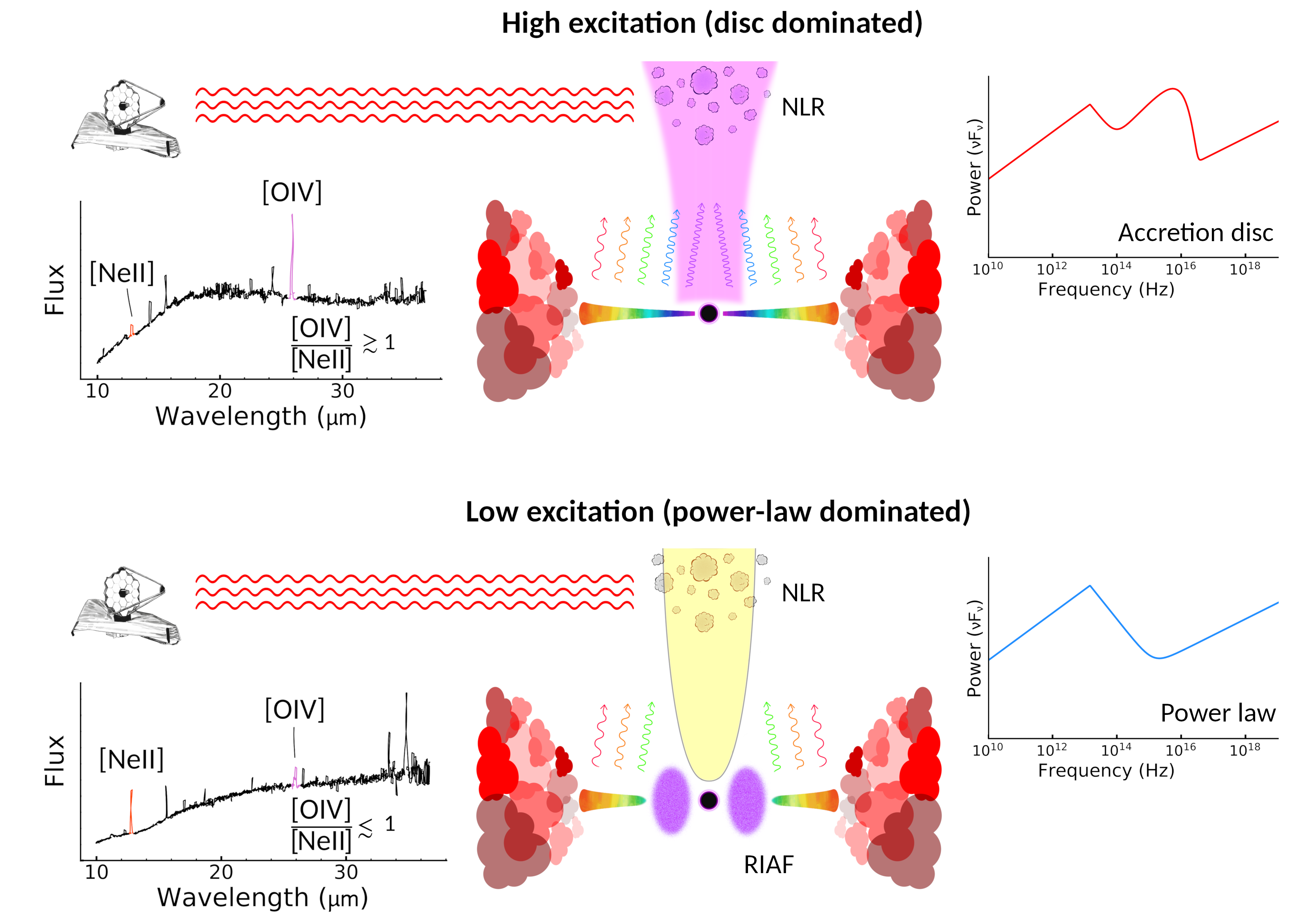}
    \caption{Sketch of a disc dominated source (top), where the gas in the narrow-line region (NLR) is illuminated by the strong UV continuum from the disc (e.g. as shown by the red solid line in the upper right panel). The latter ionises the gas producing high-excitation species such as [\ion{O}{iv}]$_{25.9}$, whose emission lines become prominent in the IR spectrum (in violet; template spectrum from NGC\,1068) when compared to low-excitation lines such as [\ion{Ne}{ii}]$_{12.8}$ (in orange). Low excitation sources (bottom) are typically dominated by the power-law continuum (synchrotron / inverse-Compton), with some or negligible contribution from a cold accretion disc and a radiatively inefficient accretion flow or RIAF (e.g. as shown by the blue solid line in the lower right panel). The lack of a strong UV continuum results in relatively faint emission lines from high-excitation gas (in violet; template spectrum from NGC\,1052), while low-excitation lines become much brighter (in orange).}\label{fig_sketch1} 
\end{figure*}

\subsection{Sensing the primary ionising continuum}\label{intro_sensing}
Understanding the nature of the ionising continuum has been a long-lasting issue in the AGN field \citep{1975ApJ...201...26B,1978Natur.272..706S,1982ApJ...254...22M,1987ApJ...323..456M,1994ApJS...95....1E,2012MNRAS.420.1848D}. Seyfert 1 nuclei tend to show a steep continuum rise shortwards of $\lesssim 1\, \rm{\micron}$, known as the featureless blue continuum \citep{1978ApJ...223...56K,1999PASP..111....1K}, which is also partially present in some Seyfert 2 nuclei \citep{1995ApJ...452..549H,2000ApJ...535..721C}. This extreme UV continuum is expected to be dominated by the accretion disc emission. For instance a (standard) accretion disc in a BH of $10^6$--$10^9\, \rm{M_\odot}$ accreting at $\sim 0.1$\,\ledd\ radiates most of its energy in the $60$--$1900$\,\AA \ range \citep{Shakura1973}. However, this emission is heavily absorbed blueward of $\sim 900$\,\AA\ due to the aforementioned high hydrogen opacity. The arrival of the \textit{Infrared Space Observatory} (\textit{ISO}; \citealt{1996A&A...315L..27K}) opened a new spectral window ($2.5$--$200\, \rm{\micron}$), which is particularly rich in ionic transitions, including the high excitation and coronal lines that respond to the most energetic part of the ionising UV continuum. Lines from ionic species above the double ionisation edge of helium ($\gtrsim 54.4\, \rm{eV}$) cannot be produced efficiently by young massive stars in starbursts, since the stellar spectra quell sharply above this limit \citep[e.g.][]{2017PASA...34...58E}. Thus, high excitation fine-structure IR lines such as [\ion{O}{iv}]$_{\rm 25.9 \mu m}$ (hereafter [\ion{O}{iv}]$_{25.9}$) and [\ion{Ne}{v}]$_{\rm 14.3, 24.3 \mu m}$ (hereafter [\ion{Ne}{v}]$_{14.3,24.3}$) should be (largely) obscuration-free tracers of the BH accretion disc \citep{1998ApJ...498..579G,2008ApJ...682...94M}. Using this argument, \citet{1999ApJ...512..204A,2000ApJ...536..710A} combined all the UV, optical, and IR lines from the spectra of NGC\,4151, Circinus, and NGC\,1068 to derive the best-fitting photo-ionising continuum that reproduces the observed line fluxes in each case. The results are consistent with ionisation by a disc emission in Circinus, while the cases of NGC\,1068 and NGC\,4151 are still inconclusive due to the presence of abrupt absorption features in the reconstructed continuum. The improved sensitivity of the \textit{Spitzer Space Telescope} \citep{2004ApJS..154....1W} allowed \citet{2011ApJ...738....6M} to increase the sample up to $\sim 70$ Seyfert galaxies and apply a similar analysis using the [\ion{O}{iv}]$_{25.9}$, [\ion{Ne}{iii}]$_{\rm 15.6 \mu m}$ (hereafter [\ion{Ne}{iii}]$_{15.6}$), and [\ion{Ne}{v}]$_{14.3, 24.3}$ high-excitation lines. The results obtained on the shape of the ionising continuum were also ambiguous, and the presence of a disc peaking at $\sim 700$--$1000$\,\AA \ could not be favoured against a simpler power-law continuum.


\section{The Luminosity-Excitation Diagram}\label{method}

A typical representation of the HLD compares the evolution of two observables: the luminosity of the source and the relative disc contribution to the total radiative output (Section\,\ref{intro}). This is convenient for BHXBs, where the emission from the main components --\,disc and corona\,-- peaks in the X-ray regime and the accretion state can be estimated by solely using high-energy observations. However, a number of major problems hamper a direct application of this approach to supermassive BHs: 
\begin{itemize}

\item To follow the evolution of single objects is unattainable due to the long characteristic time-scales involved. Therefore, state transitions in AGN --\,if present\,-- are expected to be a rare phenomenon.
\item The above-discussed challenges to study (in a direct way) the emission of the (UV-emitting) accretion disc, particularly at intermediate-to-low luminosities. 
\item BH masses in AGN typically show a large dispersion ($\sim 10^6$--$10^9\, \rm{M_\odot}$). Thus, an order-of-magnitude estimation is (at the very least) required to properly scale the total luminosity to \ledd\ when different objects are compared.
\end{itemize}

In this study we present the Luminosity-Excitation Diagram (LED), an extension of the HLD for AGN. We take advantage of the mid-IR fine-structure lines to probe the relative contribution of the disc, allowing us to investigate the existence of different accretion states in supermassive BHs. This could be expected from extrapolating the observed behaviour of BHXBs (see also simulation-based predictions by \citealt{2011MNRAS.413.2259S}), and would provide an appealing explanation for the changing-look AGN phenomenology \citep{2019ApJ...883...76R,2019arXiv190904676R}. A workaround method to infer the relative contribution of the accretion disc is proposed in Section\,\ref{LyH}. This is subsequently applied to a sample of nearby AGN with \textit{Spitzer} spectroscopy and BH mass estimates (Section\,\ref{sample}).


\begin{table*}
    \footnotesize
    \setlength{\tabcolsep}{2.pt}
	\centering
	\caption{Sample of Seyfert and LINER galaxies compiled for this study. The columns correspond to the source name, the right ascension and declination, the redshift, the distance adopted, spectral type (Sy1: Seyfert 1; Sy1h: Seyfert 1 with hidden broad lines; Sy2: Seyfert 2; LIN: LINER), BH mass, luminosity in units of \ledd, integrated line fluxes of [\ion{O}{iv}]$_{25.9}$ and [\ion{Ne}{ii}]$_{12.8}$, and radio loudness defined as the flux continuum ratio between $8.4\, \rm{GHz}$ and $2$--$10\, \rm{keV}$. Redshift-independent distances are used when available (see references in column RefD). Alternatively, distances are based on a flat-$\Lambda$CDM cosmology ($H_0 = 73\, \rm{km\,s^{-1}\,Mpc^{-1}}$, $\Omega_{\rm m} = 0.27$, $\Omega_\Lambda = 0.73$). BH masses based on dynamical estimates or reverberation mapping are used when available. Alternatively, the $\rm M_{BH}$--$\sigma_*$ relation from \citet{2013ARA&A..51..511K} was adopted, using velocity dispersion measurements compiled from the literature (see column RefM for references on $\rm M_{BH}$ or $\sigma_*$). The complete table is available in the online version of this paper.}
	\label{tab_sample}
	\resizebox{\textwidth}{!}{
	\begin{threeparttable}[b]
	\begin{tabular}{lccccccccccccc} 
		\bf Name & \bf R.A. (J2000) & \bf Dec. (J2000) & \bf z & \bf D & \bf RefD & \bf Type & \bf log(M$_{\mathbfss{\rm BH}}$) & \bf RefM & \bf log(L) &\bf [\ion{O}{iv}]$_{\rm 25.9 \mu m}$ & \bf [\ion{Ne}{ii}]$_{\rm 12.8 \mu m}$ & \bf PAH $11.3\, \rm{\micron}$ & \bf log$\left( \dfrac{F_{\mathbfss{\rm rad}}}{F_{\mathbfss{\rm X}}} \right)$ \\
		& (hh:mm:ss) & (dd:mm:ss) & & (Mpc) & & & (M$_\odot$) & & (L$_{\rm Edd}$) & ($10^{-17}\, \rm{W\,m^{-2}}$) & ($10^{-17}\, \rm{W\,m^{-2}}$) & ($10^{-17}\, \rm{W\,m^{-2}}$) & \\
		\hline
		NGC\,1052 & 02:41:04.80 & -08:15:21.0 & 0.005037 & 18.0 & 1 & LIN & 8.56 & 6 & -3.71 & $2.1 \pm 0.5$ & $21.0 \pm 0.3$ & $20 \pm 4$ & 6.40 \\
		NGC\,1068 & 02:42:40.71 & -00:00:48.0 & 0.003793 & 10.1 & 2 & Sy1h & 6.92 & 7 & -0.54 & $2030 \pm 30$ & $460 \pm 10$ & $0.11 \pm 0.01$ & 3.12 \\
		3C\,120 & 04:33:11.09 & +05:21:15.7 & 0.03301 & 139 & -- & Sy1 & 7.75 & 8 & -0.36 & $123.0 \pm 0.8$ & $7.8 \pm 0.6$ & $33 \pm 2$ & 5.93 \\
		UGC\,5101 & 09:35:51.60 & +61:21:11.0 & 0.039367 & 167 & -- & LIN & 8.37 & 9 & -1.17 & $6 \pm 1$ & $55 \pm 3$ & $460 \pm 10$ & 6.36 \\
		NGC\,4151 & 12:10:32.58 & +39:24:21.1 & 0.003319 & 16.2 & 3 & Sy1 & 7.56 & 8 & -1.58 & $244 \pm 9$ & $134 \pm 6$ & $50 \pm 10$ & 3.20 \\
		3C\,273 & 12:29:06.70 & +02:03:08.6 & 0.158339 & 728 & -- & Sy1 & 8.84 & 8 & -1.15 & $8.5 \pm 0.6$ & $1.1 \pm 0.2$ & $< 10 $ & 6.49 \\
		M87 & 12:30:49.43 & +12:23:27.9 & 0.004283 & 16.1 & 4 & LIN & 9.79 & 10 & -5.39 & $2.4 \pm 0.4$ & $7.8 \pm 0.1$ & $< 5$ & 7.65 \\
		Centaurus\,A & 13:25:27.65 & -43:01:09.2 & 0.001825 & 3.42 & 5 & Sy2 & 7.76 & 11 & -3.21 & $129 \pm 7$ & $190 \pm 20$ & $560 \pm 50$ & 4.66 \\
		\hline
	\end{tabular}
	\begin{tablenotes}
	\item 1) \citealt{2003ApJ...583..712J}; 2) \citealt{2011A&A...532A.104N}; 3) \citealt{2007A&A...465...71T}; 4) \citealt{2000ApJ...529..745F}; 5) \citealt{2007ApJ...654..186F}; 6) \citealt{2014A&A...570A..13M}; 7) \citealt{2002A&A...395L..21H}; 8) \citealt{2015PASP..127...67B}; 9) \citealt{2011ApJ...740...94D}; 10) \citealt{2011ApJ...729..119G}; 11) \citealt{2009ApJ...704L..34C}.
    \end{tablenotes}
\end{threeparttable}
}
\end{table*}

\subsection{A line ratio to probe the Lyman continuum}\label{LyH}
Fine-structure lines in nebulae are usually produced by the radiative decay of forbidden transitions from atomic and ionic levels that are collisionally-excited by free electrons in a thermal plasma. Bright emission lines from several species are typically present in the spectra of AGN, associated with the narrow-line region, where the electron densities are low enough ($n_{\rm e} \sim 10^{2-5}\, \rm{cm^{-3}}$) to avoid collisional de-excitation. The line intensities respond directly to the physical conditions of the emitting gas, and thus have been widely used in several fields as thorough diagnostics of, e.g. the gas density, temperature, and ionisation fraction \citep{2006agna.book.....O}. 

In particular, the mid-IR spectrum contains a unique suite of fine-structure lines that are virtually unaffected by dust extinction --\,a serious issue in studies of galactic nuclei\,-- and cover a vast range of physical conditions \citep{1992ApJ...399..504S}. The energy levels of these transitions are close to the ground state, which makes them fairly insensitive to the electron temperature given that (unlike for the optical lines) the involved levels become conveniently populated even in a warm medium ($T_{\rm e} \gtrsim 1000\, \rm{K}$). In addition, the different ionic species available cover a wide range of ionisation potentials (from a few $\rm{eV}$ to $> 100\, \rm{eV}$ for the coronal lines). These characteristics distinguish mid-IR lines as promising tools to sense the accretion disc emission in AGN (see Section\,\ref{intro_sensing}).

In order to probe the slope of the Lyman continuum we will use the following diagnostic, hereafter referred as the Lyman hardness:
\begin{equation}\label{eq_LyH}
    \rm{LyH_{IR}} = \frac{\rm [\ion{Ne}{ii}]_{12.8}}{\rm [\ion{Ne}{ii}]_{12.8} + [\ion{O}{iv}]_{25.9}} 
\end{equation}
where the sum of the two lines in the denominator is adopted as a proxy for the total luminosity of the ionising source. This takes into account the low-excitation gas traced by [\ion{Ne}{ii}] at $12.8\, \rm{\micron}$ (ionisation potential: $21.6\, \rm{eV}$; hereafter [\ion{Ne}{ii}]$_{12.8}$), sensitive to the $\lambda \lesssim 575$\,\AA \ continuum, and the high-excitation gas traced by [\ion{O}{iv}] at $25.9\, \rm{\micron}$ ($54.9\, \rm{eV}$), which responds to the continuum below $\sim 226$\,\AA. LyH$_{IR}$ probes the spectral slope in the extreme UV, and thus to the possible contribution of the accretion disc emission to the ionising radiation. This is illustrated in the sketch shown in Fig.\,\ref{fig_sketch1} (top panel), where the ionisation of the narrow-line region gas by the strong UV continuum from the disc (red line in the upper-right continuum spectrum) produces a prominent emission in the [\ion{O}{iv}]$_{25.9}$ line. Contrastingly, colder discs in putative hard-state AGN should cause a dearth of extreme UV photons (blue line in the lower-right spectrum), and therefore lower [\ion{O}{iv}]$_{25.9}$ intensities and higher LyH$_{IR}$ values. This is the basis of our working hypothesis, that is, a disc-dominated system will show [\ion{O}{iv}]$_{25.9}$/[\ion{Ne}{ii}]$_{12.8} \gtrsim 1$ and therefore a low LyH$_{IR}$ value ($\lesssim 0.5$), while a power-law continuum (i.e. no disc) would result in [\ion{O}{iv}]$_{25.9}$/[\ion{Ne}{ii}]$_{12.8} \lesssim 1$ and consequently LyH$_{IR} \gtrsim 0.5$, approaching LyH$_{IR}\sim 1$ when the [\ion{O}{iv}]$_ {25.9}$ flux becomes negligible. In this context, the Lyman hardness can be regarded as LyH$_{IR} \sim 1 - L_{\rm disc} / L_{\rm total}$. This hypothesis and the expected parameter range will be further explored using photo-ionisation models in Section\,\ref{photmod}. The selection of [\ion{O}{iv}]$_{25.9}$ as a genuine tracer of AGN accretion is the key element in our analysis, since it is essentially the only high-excitation line above the helium edge at $54.4 \rm{eV}$ that enables to include low-luminosity AGN. We note that these objects are characterised by ionising continua that are generally too weak to produce prominent higher excitation lines (e.g. [\ion{Ne}{v}]$_{14.3,24.3}$).

Given the typical size of the narrow-line region (from $\sim 10$ to hundreds of parsecs), the nebular lines are not expected to react quickly to possible changes in the continuum emission. This is the case of NGC\,4151, where the flux of the [\ion{O}{iii}] line at $5007$\,\AA \ has been stable for decades despite the continuum variability and spectral type transitions \citep{1983ApJ...271..564A,1997A&A...324..904M}. However, very compact (from one light-year to a few parsecs), and mildly variable (from months to tens of years) narrow-line regions have been observed in Seyfert galaxies (e.g. 3C\,390.3 in \citealt{1995AJ....109.2355Z}; NGC\,5548 in \citealt{2013ApJ...779..109P}; Mrk\,590 in \citealt{2014ApJ...796..134D}). Transitions from high-ionisation species are expected to show a higher variability, particularly in the case of coronal lines, since they are formed in the innermost part of the narrow-line region \citep[e.g.][]{2019ApJ...874...44Y,2019ApJ...883...31F}.

\begin{figure}
	\includegraphics[width=\columnwidth]{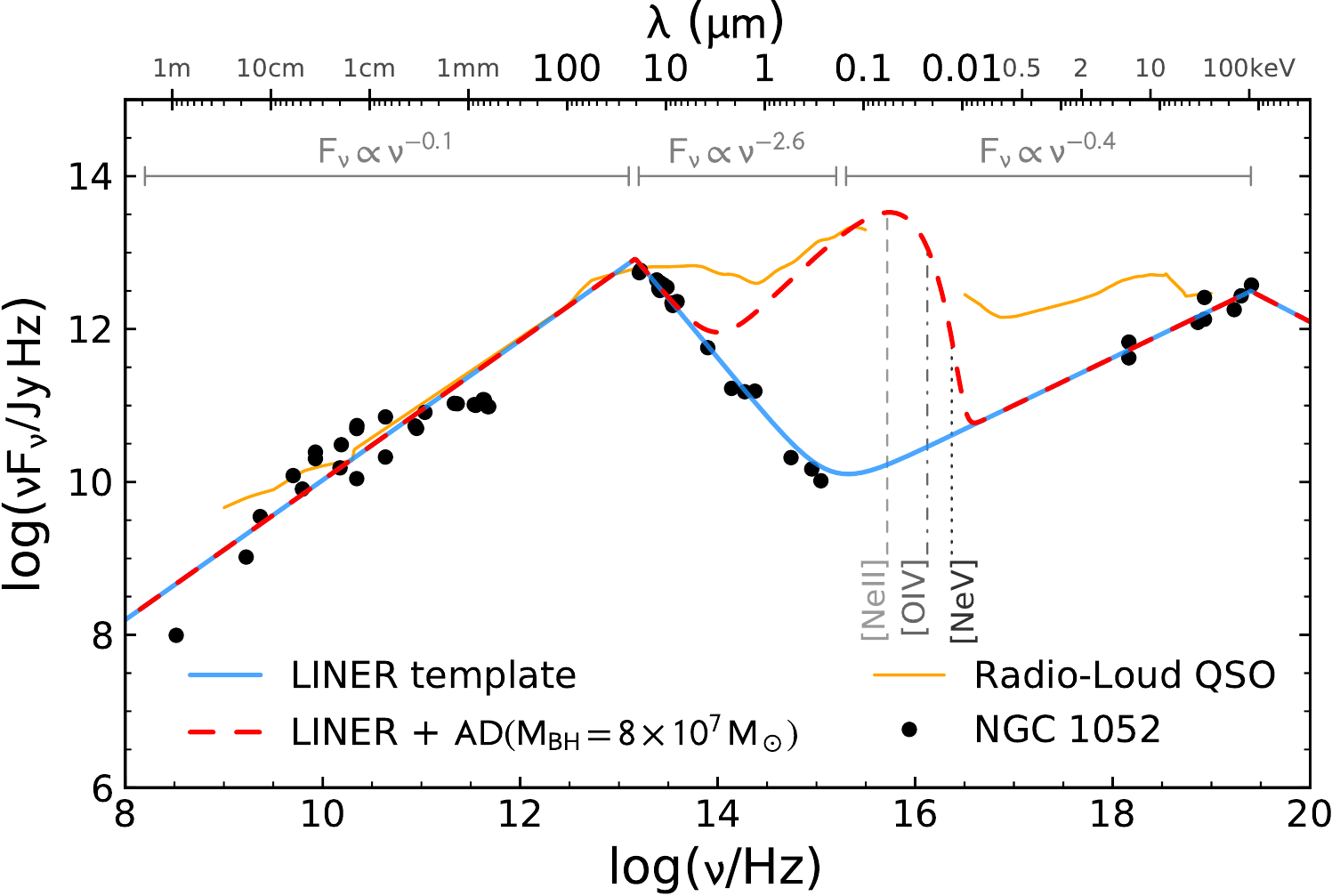}
    \caption{Nuclear spectral energy distribution of NGC\,1052 based on high-angular resolution data ($\lesssim 0\farcs4 \sim 30\, \rm{pc}$, black dots; \citealt{2019MNRAS.485.5377F}). Input spectra used for \textsc{Cloudy} photo-ionisation simulations are also shown: a piecewise power-law fit to the flux distribution of NGC\,1052 (blue solid line) and the same model plus a standard accretion disc (red dashed line; \citealt{Shakura1973}). The accretion disc contribution has been scaled to mimic the radio-loud quasar template (orange solid line; \citealt{1994ApJS...95....1E}). The ionisation potentials of \textsc{Ne}$^+$ ($21.6\, \rm{eV}$), O$^{3+}$ ($54.9\, \rm{eV}$), and \textsc{Ne}$^{4+}$ ($97.1\, \rm{eV}$) are indicated as vertical dashed, dot-dashed, and dotted lines, respectively.}
    \label{fig_input}
\end{figure}

\subsection{The sample}\label{sample}
Our sample comprises mid-IR line fluxes of 154 AGN obtained with the \textit{Spitzer}/Infrared Spectrograph (IRS) using the high-spectral resolution (HR) mode ($10$--$37\, \rm{\micron}$, $R \sim 600$). These include Seyfert galaxies from the $12$ micron sample of galaxies \citep{1993ApJS...89....1R}, whose mid-IR line fluxes were measured by \citet{2008ApJ...676..836T,2010ApJ...709.1257T}. To cover the low-luminosity domain we incorporated the AGN spectroscopic database in \mbox{\citet{2016ApJS..226...19F}} plus \textit{Spitzer}/IRS-HR measurements of LINERs from \citet{2009ApJ...691.1501D} and \citet{2009MNRAS.398.1165G}. Finally, five AGN observed with the Short Wavelength Spectrometer ($2.4$--$45\, \rm{\micron}$, $R \sim 1000$--$2000$; \citealt{1996A&A...315L..49D}) onboard \textit{ISO} were added \citep{2002A&A...393..821S,2003A&A...403..829V,2004A&A...414..825S}, for a total sample of 167 local Universe AGN.

Our aim is to populate the LED covering a wide range in AGN luminosities, from Seyfert-like nuclei where the presence of the disc is expected to dominate the extreme UV continuum, to low-luminosity AGN, where is likely that the disc is significantly fainter. We note, nonetheless, that this is not a complete sample. While luminosity and excitation conditions span a very wide range, the density of sources in the different regions of the diagram could be affected by selection biases. Nevertheless, the majority of the AGN used in this work (97 out of 167) are part of the 12 micron sample \citep{1993ApJS...89....1R}, which is a flux-limited sample of galaxies at $12\, \rm{\micron}$ selected from the IRAS Faint Source Catalogue, complete down to $0.3\, \rm{Jy}$ \citep{1989ApJ...342...83S,1993ApJS...89....1R}. Thus, the main results of this study are not likely affected by strong biases.

\begin{figure*}
	\includegraphics[width=0.9\textwidth]{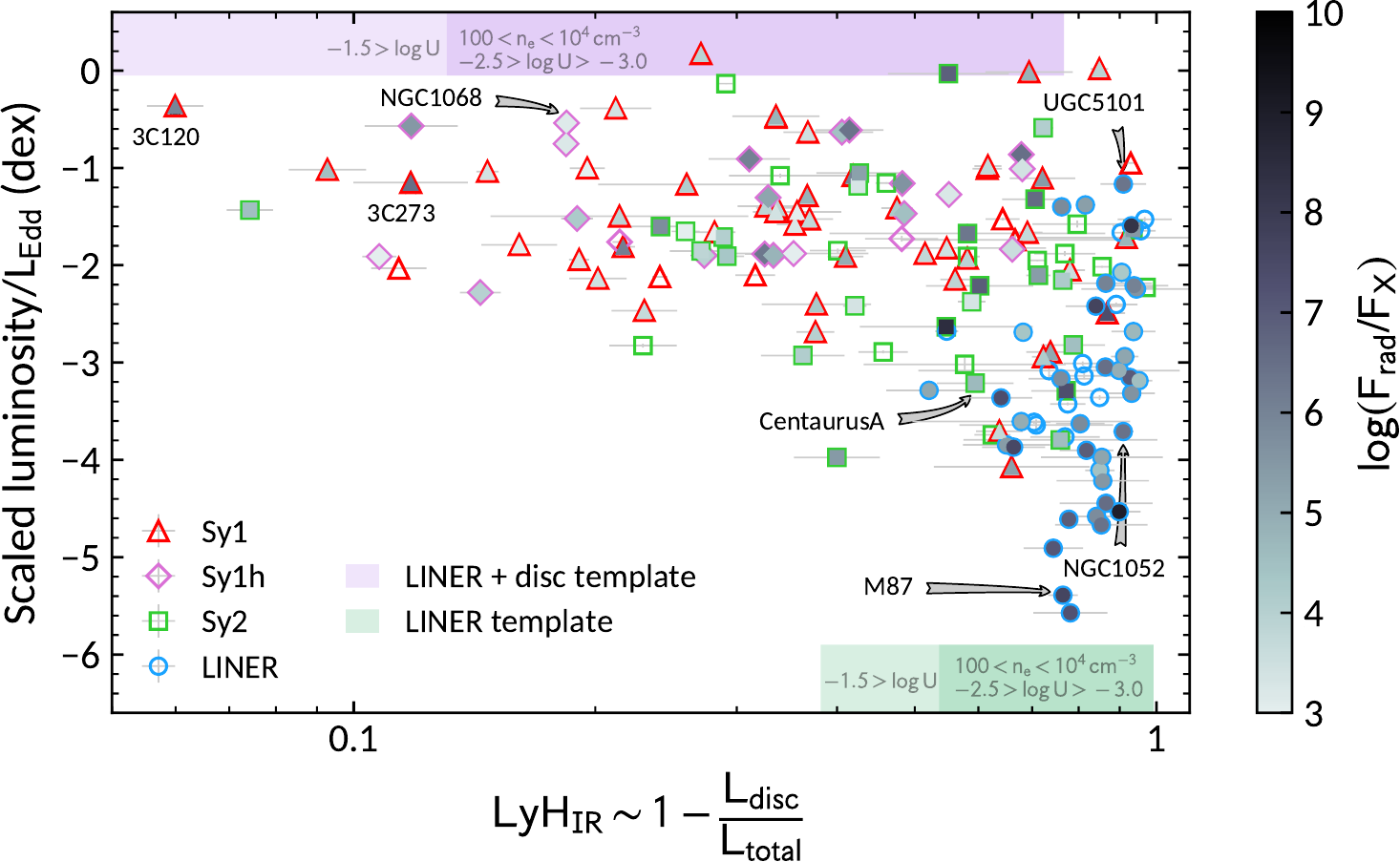}
    \caption{The luminosity-excitation diagram (LED) for AGN represents the excitation traced by mid-IR fine-structure lines [LyH$_{IR} =$\,[\ion{Ne}{ii}]$_{12.8}$/([\ion{Ne}{ii}]$_{12.8}$ + [\ion{O}{iv}]$_{25.9}$)] against the total  line luminosity (re-scaled; see text) expressed in units of \ledd \ ($L_{\rm lines} = L_{\rm [NeII]12.8} + L_{[OIV]25.9}$). The AGN are grouped following their optical spectral classification, i.e. Sy1 (red triangles), Sy1h (type 1 with hidden broad lines; violet diamonds), Sy2 (green squares), and LINER (blue circles).  The colour-shaded areas represent the range in excitation covered by our \textsc{Cloudy} photo-ionisation models for the LINER template (in green) and the LINER plus disc template (in purple). Darker areas correspond to models with $100 < n_{\rm e} < 10^4\, \rm{cm^{-3}}$ and $-2.5 > \log U > -3.0$, extended to $-1.5 > \log U$ for the light-coloured areas. Filled symbols are coloured according to their radio ($8.4\, \rm{GHz}$) to X-ray ($2$--$10\, \rm{keV}$) continuum ratio. Objects for which this ratio could no be computed are depicted by empty symbols.}\label{fig_led}
\end{figure*}

BH masses are required to scale luminosities with the corresponding \ledd. The best determinations are based on either dynamical models of the central regions of the galaxies (compiled by \citealt{2013ARA&A..51..511K}) or reverberation mapping techniques \citep{2015PASP..127...67B}. However, these are only available for $\sim 30\%$ of the sample. BH masses for the remaining sources were derived from the $M_{\rm BH}$--$\sigma$ scaling relation \citep{1998AJ....115.2285M,2000ApJ...539L...9F}, using the stellar velocity dispersion in the host galaxy bulge (from the HyperLEDA database; \citealt{2014A&A...570A..13M}). The formal expression of the $M_{\rm BH}$--$\sigma$ relation adopted in this work is given in \citet{2013ARA&A..51..511K}, and it was obtained from dynamical measurements of BH masses in 87 nearby galaxies. This calibration has an intrinsic scatter of $\sim 0.3\, \rm{dex}$, which is folded into our BH mass estimates. Nevertheless, this uncertainty is significantly lower than the $\gtrsim 2$--$3$ orders of magnitude dispersion expected if luminosities are not scaled by \ledd.

The entire sample (167 AGN) is detected in both [\ion{O}{iv}]$_{25.9}$ and [\ion{Ne}{ii}]$_{12.8}$, and has a BH mass estimate. This includes 55 Seyfert 1 nuclei (Sy1; red triangles in Fig.\,\ref{fig_led}), 22 Seyfert 1 with hidden-broad lines (Sy1h; violet diamonds), 39 Seyfert 2 (Sy2; green squares), and 51 LINERs (blue circles). As a result of dust obscuration, Sy1hs do not have an optical broad line region (unlike Sy1s), which is instead revealed in either the polarised or the IR spectrum. The adopted AGN spectral classification is based on the ratios of optical emission lines provided by \citet{2010A&A...518A..10V}, while the different AGN classes have been grouped following the approach by \citet{2010ApJ...709.1257T}. We note that the spectral types were compiled by \citet{2010A&A...518A..10V} from different studies, and therefore the spectral classification quoted in Table\,\ref{tab_sample} is not based on a single homogeneous dataset. Thus, some sources could be re-classified in the future if new high-quality spectra are acquired (e.g. deep spectro-polarimetric observations).

As in Eq.\,\ref{eq_LyH}, we adopt the total line flux from [\ion{Ne}{ii}]$_{12.8}$ and [\ion{O}{iv}]$_{25.9}$ as a proxy for the luminosity of the ionising source, using redshift-independent distances given by NED\footnote{The NASA/IPAC Extragalactic Database (NED) is funded by the National Aeronautics and Space Administration and operated by the California Institute of Technology.} when available, or alternatively distances based on a flat-$\Lambda$CDM cosmology ($H_0 = 73\, \rm{km\,s^{-1}\,Mpc^{-1}}$, $\Omega_{\rm m} = 0.27$, $\Omega_\Lambda = 0.73$). Assuming that the highest Eddington ratios found in our sample are close to $\sim 1$, we estimate a normalised line luminosity of $L^{\prime}_{\rm lines} = 10^{-3} \times (L_{[\ion{O}{iv}]25.9} + L_{[\ion{Ne}{ii}]12.8})$. This is in agreement with the available estimates for well characterised sources. For instance, in M87 the normalised line luminosity, $\log(L'_{\rm line} / L_{\rm Edd}) = -5.4$ (Fig.\,\ref{fig_led}), coincides with the Eddington ratio obtained from the integrated spectral energy distribution, $\log(L_{\rm bol}/ L_{\rm Edd}) \sim -5.4$ \citep{1996MNRAS.283L.111R,2016MNRAS.457.3801P}. Additional examples are NGC\,1052 ($-3.7$ vs. $-3.4$ from \citealt{2019MNRAS.485.5377F}), Centaurus\,A ($-2.7$ vs. $-2.4$ from \citealt{2011A&A...531A..70B}), NGC\,1068 ($-0.5$ vs. $-0.2$ from \citealt{2020MNRAS.492.3872Z}), and 3C\,120 ($-0.4$ vs. $-0.5$ from \citealt{2016MNRAS.458.2454L,2018ApJ...856..120R}).

A complete table for the sample of galaxies can be found as part of the online materials (see Table\,\ref{tab_sample}), including the adopted distances, BH masses, luminosities and compiled line fluxes. 

\begin{figure*}
	\includegraphics[width=0.9\textwidth]{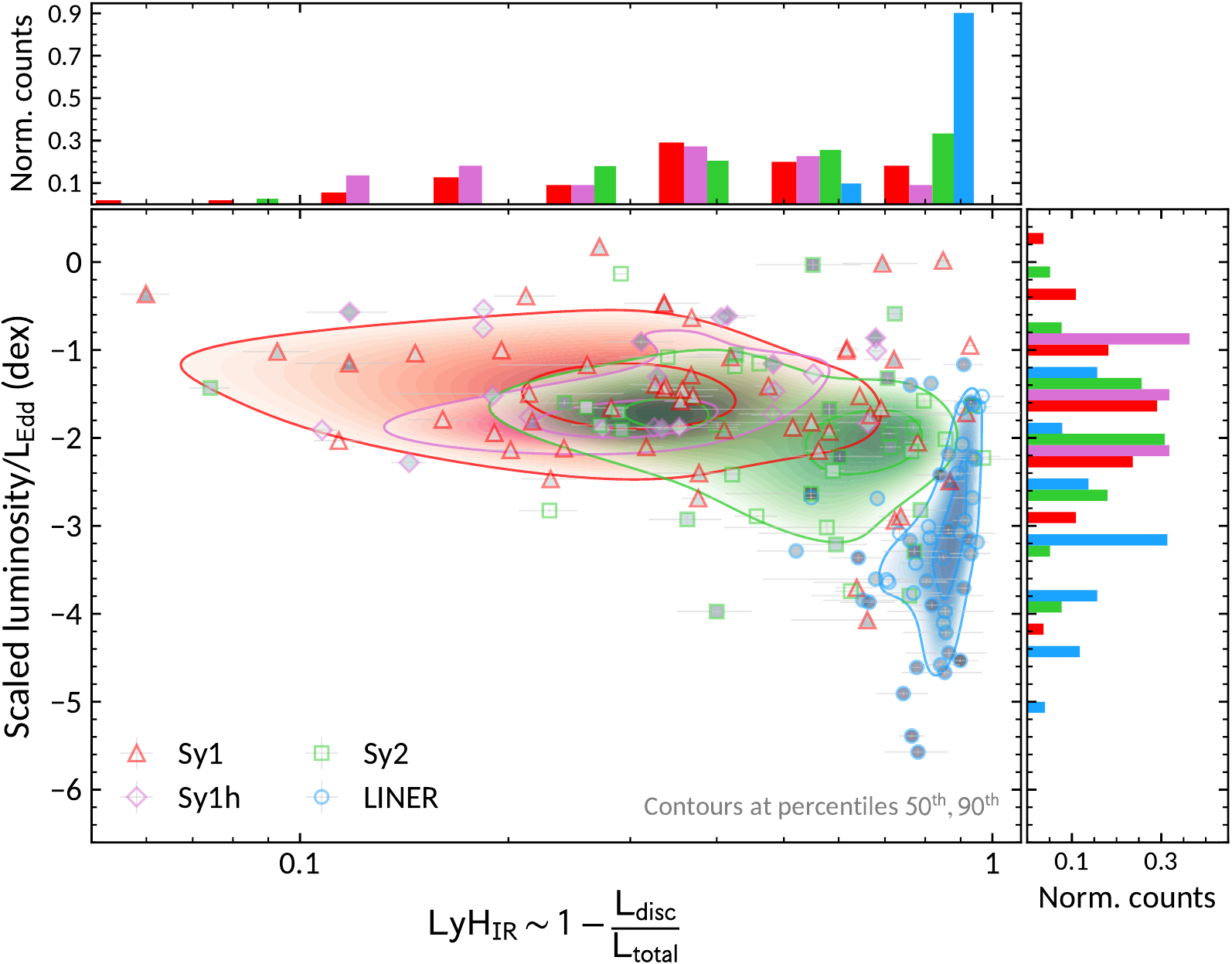}
    \caption{Luminosity-excitation diagram (LED) for AGN based on mid-IR fine-structure lines (LyH$_{IR} =$\,[\ion{Ne}{ii}]$_{12.8}$/([\ion{Ne}{ii}]$_{12.8}$ + [\ion{O}{iv}]$_{25.9}$)) and the scaled line luminosity in \ledd\ units. AGN are grouped by their optical spectral classification (symbols as per Fig.\,\ref{fig_led}). For each spectral class, the probability density distribution has been obtained using a Gaussian kernel density estimation \citep{1986desd.book.....S} on the corresponding scatter plot. Solid contours correspond to percentiles $50^{\rm th}$ and $90^{\rm th}$. The marginal distributions in luminosity and excitation for the different AGN types are shown on the side panels in normalised counts.}\label{fig_dens}
\end{figure*}

\subsection{Building the LED diagram}\label{led}
The resulting diagram is shown in Fig.\,\ref{fig_led}, where different symbols correspond to the canonical AGN spectral types. Luminous Seyfert nuclei with $\log(L^{\prime}_{\rm lines}/ \rm{L_{Edd}}) \gtrsim -3$ are distributed over a wide range along the excitation axis (0.1 $\lesssim \rm{LyH_{IR}} \leq 1$). By contrast, LINERs are remarkably absent in the left (high-excitation) side of the diagram. The latter populate a narrow fringe at very low excitations ($\rm{LyH_{IR}} \gtrsim 0.6$) reaching two to three orders of magnitude lower luminosities when compared to Seyferts. Nonetheless, a few LINERs are found among the brightest sources, and some Seyferts show sub-Eddington luminosities. The location of the LINERs in the low excitation side of the LED indicates a dearth of highly energetic photons in these nuclei, which are not able to produce a significant amount of [\ion{O}{iv}]$_{25.9}$ when compared to most of the Seyfert nuclei. 

From the scatter plot in Fig.\,\ref{fig_led} we derived the probability density distribution for each of the AGN (optically-defined) types. We applied a kernel density estimation method \citep{1986desd.book.....S}, adopting a Gaussian kernel, which returns a probability distribution that matches the discrete data sample. The result is shown in Fig.\,\ref{fig_dens}, where contours for each AGN distribution correspond to the $50^{\rm th}$ and $90^{\rm th}$ percentiles. The density representation and the marginal distributions in Fig.\,\ref{fig_dens} confirm the narrow range covered by the LINERs in the excitation axis, with most of them having very low luminosities ($10^{-3}$--$10^{-4}\, \rm{L_{Edd}}$), although they reach up to $\sim 0.1\, \rm{L_{Edd}}$. The loci of type 1 nuclei (Sy1s and Sy1hs) overlap, with most of the sources in the high-excitation, high-luminosity part of the diagram. Type 2 Seyferts, on the other hand, show a bimodal distribution with two maxima along the excitation axis. A Kolmogorov-Smirnov test on the distribution of $\rm{LyH_{IR}}$ values gives a probability lower than $1\%$ for the Sy2 nuclei to come from the same parent sample as the Sy1 sample. On the other hand, the comparison between Sy1s and Sy1hs returns a probability of $60\%$.

\subsection{Radio loudness}
Alongside with the likely variable disc contribution, a key aspect in the study of accreting BHs is the high radio loudness associated with the hard state, due to the prevalence of the jet during this phase \citep[e.g.][]{Fender2004}. In order to characterise the radio loudness in our sample we use the continuum flux ratio of radio waves to X-rays ($F_{\rm rad}/F_{\rm X}$, with $F_{\rm rad}$ and $F_{\rm X}$ corresponding to continuum flux densities at $8.4\, \rm{GHz}$ and $2$--$10\, \rm{keV}$, respectively). The hard X-ray emission in AGN (as well as in BHXBs) is typically dominated by upscattered seed photons from the disc by inverse Compton cooling of the BH corona \citep[e.g.][]{1991ApJ...380L..51H,Gallo2003,2019ApJ...875..133P}, and thus it is usually adopted as a tracer of the accretion luminosity \citep[e.g.][]{2019MNRAS.tmp.3152S}. This ratio is expected to peak during the hard state due to stronger radio emission from the jet, and the dimming of the X-ray luminosity. We note that this radio loudness definition differs from the classical radio-to-optical ratio in AGN, which is usually dominated by the stellar light in the optical (unless $L \sim L_{\rm Edd}$). 

An extensive search in the literature was performed to compile nuclear X-ray luminosities ($2$--$10\, \rm{keV}$; corrected for intrinsic absorption) and radio continuum measurements for the galaxies in our sample. Radio fluxes at $8.4\, \rm{GHz}$ were used as a compromise to gather the largest possible coverage at the highest frequency, in order to amend the effects of self-synchrotron absorption. When data at $8.4\, \rm{GHz}$ were not available, we performed a linear interpolation of the flux density between the closest measurements in frequency. Because of the higher opacity, the low frequency continuum tends to escape from farther regions in the jet (e.g. \mbox{\citealt{2011Natur.477..185H}}), increasing the scatter when compared to the more transparent continuum at high-frequencies, which is emitted closer to the BH. The majority of the $8.4\, \rm{GHz}$ measurements were taken from \citet{Thean2000}, whose data set was obtained using the Very Large Array (A-configuration) that renders a typical angular resolution of $\lesssim 0.5$ arcsec, thus isolating the core emission from the extended lobes. For the remaining sources observed at lower angular resolution, we note that possible contamination from unresolved lobes should be significantly amended by the selection of a relatively high frequency ($8.4\, \rm{GHz}$), which is more sensitive to the radio core. We do not use measurements at millimetre/submillimetre wavelengths because they are less numerous and might be contaminated by the Rayleigh-Jeans tail of cold dust emission.

\subsubsection{The radio loudness of our sample}
Symbols in Fig.\,\ref{fig_led} are coloured according to the corresponding radio loudness, while sources missing either X-ray or radio measurements are represented with empty symbols. Once again LINERs stand out in the diagram due to their high radio loudness, showing typical values of $\log(F_{\rm rad}/F_{\rm X}) \gtrsim 6$ when compared to X-ray brighter radio-quiet nuclei with $\log(F_{\rm rad}/F_{\rm X}) \lesssim 5$. Overall, the low-excitation Seyferts on the right side of the LED tend to be also relatively radio loud when compared to the rest of the Seyfert population. That is, sources with a low radiative efficiency and/or a weak ionising continuum tend to show a higher activity in radio (see Section\,\ref{hard-state}).

The radio loudness values for the entire sample can be found in the online materials associated with this publication.

\section{Comparison with photo-ionisation models}\label{photmod}
To interpret our spectral diagnostics and the nature of the ionising continuum on more physical grounds, we have computed a set of photo-ionisation simulations using \textsc{Cloudy} (version C17.02; \mbox{\citealt{2017RMxAA..53..385F}}), assisted by the py\textsc{Cloudy} library \citep{2013ascl.soft04020M}. The models assume a plane-parallel geometry with constant pressure and solar abundances \citep{2009ARA&A..47..481A}, sampling a grid in both density ($n_{\rm H} = 1$--$10^6\,\rm{cm^{-3}}$, step $0.5\, \rm{dex}$) and ionisation parameter ($-3.5 < \log U < -1.0$, step $0.5\,\rm{dex}$). Regarding the primary ionising spectrum, we computed two main sets of simulations intended to reproduce the most extreme cases in the LED: \textit{i)} low-excitation nuclei with bright [\ion{Ne}{ii}]$_{12.8}$ emission and faint [\ion{O}{iv}]$_{25.9}$; \textit{ii)} the high-excitation nuclei where [\ion{O}{iv}]$_{25.9}$ can be the most prominent line in the mid-IR spectrum (e.g. NGC\,1068 and 3C\,120 in \mbox{\citealt{2010ApJ...709.1257T}}). Our aim is to make an overall comparison between the measured line ratios in Fig.\,\ref{fig_led} and the photo-ionisation expected for a LINER-like continuum, rather than developing sophisticated models for detailed analysis of individual sources \citep[e.g.][]{2015ApJ...801...42D}.

As a reference template for the ionising continuum of low-excitation LINER nuclei, we adopted a power-law piecewise fit to the flux distribution of NGC\,1052 (blue solid line in Fig.\,\ref{fig_input}). The fit relies on flux measurements obtained with a remarkably high angular resolution dataset ($\lesssim 0\farcs4 = 35\, \rm{pc}$ at the distance of NGC\,1052) and compiled by us (black dots in Fig.\,\ref{fig_input}; \citealt{2019MNRAS.485.5377F}). The choice of NGC\,1052 as a reference for low-excitation nuclei has a twofold motivation: \textit{i)} the steep power-law UV continuum found to be common among the LINER class (of which NGC\,1052 is a prototypical example; \citealt{1996ApJ...462..183H,1980A&A....87..152H}), that misses the big blue bump continuum associated with the disc emission; \textit{ii)} the superior sampling in wavelength achieved for this galaxy due to the lack of dust obscuration, which allows us to detect the near-UV continuum. Since direct measurements of the ionising continuum cannot be obtained, the template was interpolated between the bluest UV measurement at $\sim 2600$\,\AA \ and the X-ray continuum following a power law with $F_\nu \propto \nu^{-0.4}$. This approximation seems reasonable at first order from the trend suggested by the IR-to-UV continuum and the X-ray measurements in Fig.\,\ref{fig_input}, and it is similar to the approach followed by \citet{2000ApJ...532..883G}.

The width of the dark-green shaded area in the lower right side of Fig.\,\ref{fig_led} shows the range in LyH$_{IR}$ values occupied by LINER templates with densities between $100$ and $10^4\, \rm{cm^{-3}}$ and ionisation parameter $-2.5 > \log U > -3.0$. All the LINERs data points (circles) fall inside this region ($0.52 < \rm{LyH_{IR}} < 1$). This indicates that no significant excess is required above the interpolated power law (blue solid line in Fig.\,\ref{fig_input}) to describe, at first order, the nebular line excitation in LINERs. The extreme UV continuum slope of $-0.4$ is quite close to the high-energy slope of the AGN template derived by \citet{1987ApJ...323..456M} and incorporated into \textsc{Cloudy} (see fig.\,5 in \citealt{2013RMxAA..49..137F}), which scales as $\propto \nu^{-0.5}$ shortwards of $\sim 30$\,\AA. The observed LyH$_{IR}$ values in our LINER sample suggest, however, that such component extends to longer wavelengths ($\sim 600$\,\AA) for these nuclei. Increasing the ionisation parameter in our models to $\log U = -1.5$ lowers the predicted $\rm{LyH_{IR}}$ values down to $\gtrsim 0.4$ (light-grey area in Fig.\,\ref{fig_led}), while further increments in $\log U$ result in negligible changes in $\rm{LyH_{IR}}$. This means that models using the adopted LINER template cannot reach the left side of the LED in Fig.\,\ref{fig_led} due to the lack of hard ionising photons.

\subsection{Simulating the disc contribution}
If present, the accretion disc should dominate the extreme UV continuum and therefore the fine structure line ratios, in particular for those ionic species with high ionisation potential values ($\gtrsim 55\, \rm{eV}$), i.e. O$^{3+}$ in our case. In order to investigate the excitation at very low LyH$_{IR}$ values, we modified the NGC\,1052 power-law template by adding the expected contribution from a standard, geometrically thin disc \mbox{\citep{Shakura1973}}, assuming a BH mass of $8 \times 10^7\, \rm{M_\odot}$ --\,the median BH mass value in our sample\,-- accreting at $0.1\, \dot{M}_{\rm Edd}$ (red dashed template in Fig.\,\ref{fig_input}). The purpose of these simulations is to probe the impact on the line ratios of an excess above the adopted LINER template continuum. From this point of view, the thermal-like contribution of the standard disc is a valid first-order approximation, while a detailed reconstruction of the primary ionising continuum is beyond the scope of this work \citep[see][]{1999ApJ...512..204A,2000ApJ...536..710A,2011ApJ...738....6M}. The disc emission has been normalised to mimic the continuum shape of a standard radio-loud quasar template (orange line in Fig.\,\ref{fig_input}; \citealt{1994ApJS...95....1E}). As expected, the photo-ionisation simulations including the disc component predict a large increase of [\ion{O}{iv}]$_{25.9}$ over [\ion{Ne}{ii}]$_{12.8}$, shifting $\rm{LyH_{IR}}$ towards much lower values, i.e. $\sim 0.12$ for $\log U = -2.5$ at $100\, \rm{cm^{-3}}$, and $\rm{LyH_{IR}} \lesssim 0.05$ for $\log U = -1.5$ (dark and light-purple shaded areas at the top of Fig.\,\ref{fig_led}, respectively). The observed ratios of the Seyfert galaxies are reproduced by these models, supporting a significant excess in the extreme UV continuum over the flat LINER power-law template.

The accretion disc emission shifts towards lower frequencies for higher BH masses, thus reducing its capability to produce high-energy photons. This has a critical effect if one simply takes the \citet{Shakura1973} model at face value, which predicts a very steep flux drop bluewards of the disc emission peak. For instance, to assume a BH mass of $10^9\, \rm{M_\odot}$ in the (dashed red line) template in Fig.\,\ref{fig_input} would result in an emission bump that do not reach the ionisation potential of O$^{3+}$ and Ne$^{4+}$, and therefore would return LyH$_{IR} \sim 1$ even in the presence of an accretion disc. This might be the case for the 11 sources (out of 167) that have BH mass estimates above this value. However, two of these objects, Cygnus\,A and 3C\,433, show instead low LyH$_{IR}$ values of $0.2$, and $0.3$, respectively, suggesting that the drop in flux shortwards of the disc emissivity peak may be more gradual, in line with the (empirical) quasar continuum template (solid, orange line in Fig.\,\ref{fig_input}). Likewise, M87 shows a low excitation level (LyH$_{IR} = 0.8$) but it hosts one of the most massive BHs of the sample, which might prevent the detection of the accretion disc by this method. However, this disc, if present at all, should be detectable at optical/UV wavelengths, whereas it is not revealed by high angular resolution observations of the nuclear region \citep{2016MNRAS.457.3801P}. This supports the non-thermal nature for the ionising continuum (i.e. in agreement with the high LyH$_{IR}$ value).

Finally, we note that shocks are not included as a source of excitation in our simulations. We acknowledge that they might dominate the gas excitation in some exceptional cases \citep[e.g. NGC\,1386;][]{2017MNRAS.470.2845R}, but bulk velocities of the order of $\gtrsim 400\, \rm{\kms}$ would be required to produce a significant increase in the [\ion{O}{iv}]$_{25.9}$ flux \citep{2001ApJS..137...75C,2008ApJS..178...20A}, which is not likely the case for most of the galaxies in our sample. Slow shocks could affect the [\ion{Ne}{ii}]$_{12.8}$ emission, but then a high [\ion{Ne}{ii}]$_{12.8}$/[\textsc{Ne\,iii}]$_{15.6}$ ratio of about $10$--$100$ would be expected \citep[e.g.][]{2000ApJ...529..219S}, while the median value in our sample of LINERs is $2.3 \pm 1.5$.

\begin{figure*}
	\includegraphics[width=0.9\textwidth]{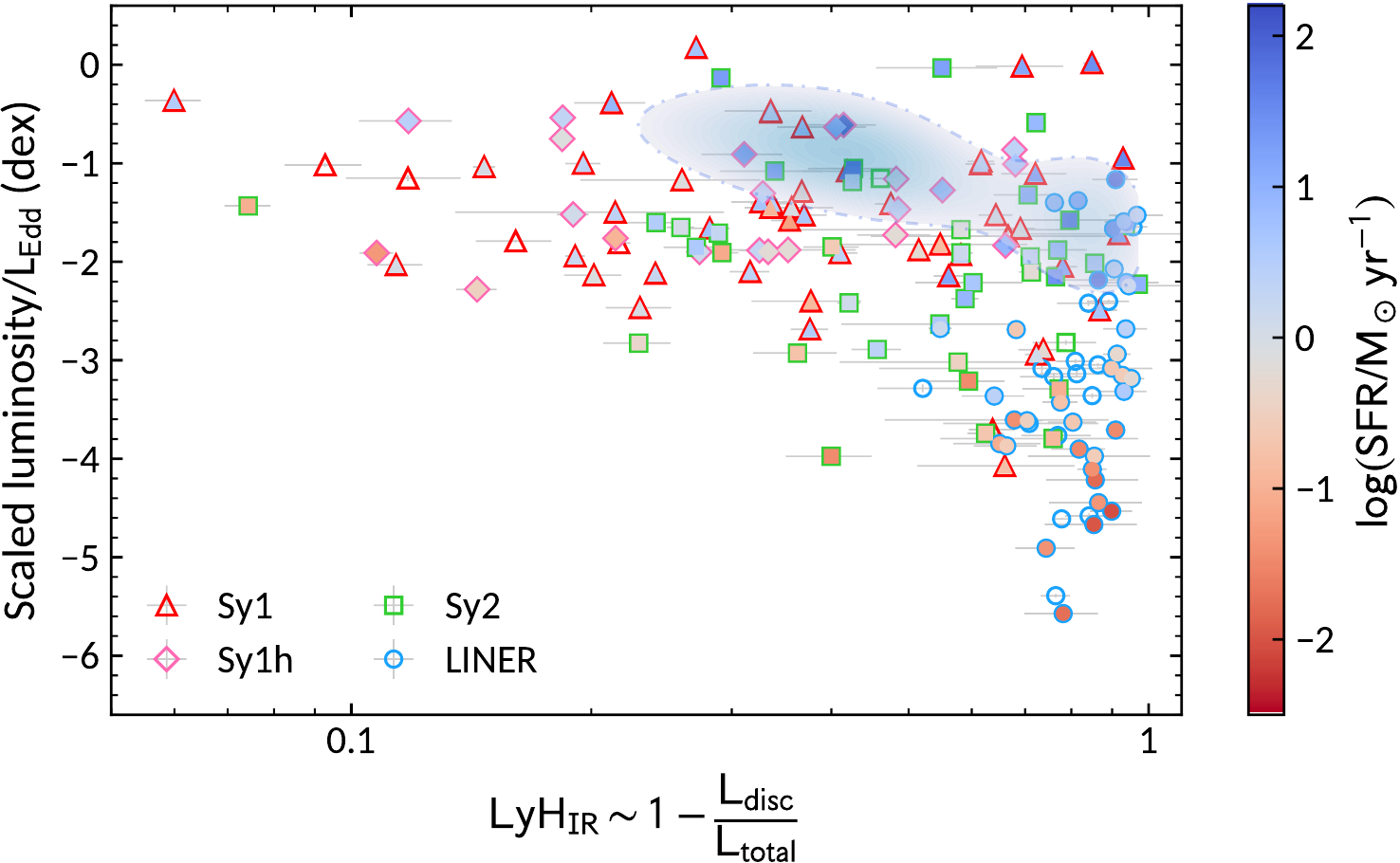}%
    \caption{Luminosity-excitation diagram (LED) for AGN based on mid-IR fine-structure lines (LyH$_{IR} =$\,[\ion{Ne}{ii}]$_{12.8}$/([\ion{Ne}{ii}]$_{12.8}$ + [\ion{O}{iv}]$_{25.9}$)) and the scaled line luminosity expressed in \ledd\ units. AGN are grouped by their optical spectral classification (symbols as per Fig.\,\ref{fig_led}). Symbols are colour-filled according to their star-formation rate derived from the PAH emission band at $11.3\, \rm{\micron}$ (see colour bar). Objects lacking a star-formation rate estimate are depicted by empty symbols. The probability density distribution for the sample, weighted by the star formation rate, is shown by the blue shaded area, with the $50^{\rm th}$ percentile contour delimited by the dot-dashed line.}\label{fig_sfr}
\end{figure*}

\section{Discussion}\label{discuss}

The aim of this paper is to investigate whether the accretion states widely observed in X-ray binaries (Section\,\ref{intro}) might be present to some extent in supermassive BHs and can be identified in the LED (Figs.\,\ref{fig_led} and \ref{fig_dens}).

\subsection{Comparison with previous works}\label{discuss_prev}
In the disc fraction luminosity diagram presented by \citet{Koerding2006} and more recently by \citet{2017A&A...603A.127S}, the X-ray luminosity is adopted as a measure of the contribution from the corona, while either the \textit{B}~band magnitude or the $2900$\,\AA \ continuum flux, respectively, are used as a proxy for the disc luminosity. In order to incorporate LINERs in this analysis (dominated by starlight in the optical), \citet{Koerding2006} use the H$\beta$ emission line flux as a proxy for the nuclear \textit{B}~band magnitude. \citet{2017A&A...603A.127S} use simultaneous near-UV and X-ray measurements to avoid additional scatter due to possible variability. Both studies find that stronger radio emission is associated with a larger X-ray to optical continuum ratio (i.e. hardness), although \citet{2017A&A...603A.127S} warn that the UV emission in low-luminosity systems is strongly affected by starlight from the host galaxy. When compared to the optical or the near-UV continuum, the LyH$_{IR}$ is arguably a more selective tracer, because it probes the peak of the expected disc contribution. This is critical in moderate-to-low luminosity AGN ($\le 10^{43}\, \rm{erg\,s^{-1}}$) but also at low BH masses ($\lesssim 10^{7}\, \rm{M_\odot}$), since the disc emission is shifted towards shorter wavelengths and its contribution to the optical continuum becomes negligible. In this regard, the inclusion of a high excitation line such as [\ion{O}{iv}]$_{25.9}$ in the diagnostics is fundamental, given that the stellar populations cannot contribute significantly to its flux.

Recently, \citet{2020arXiv200407258A} compared the luminosities of the accretion disc and the corona in luminous AGN to those of the prototypical BHXB GX\,339--4 during its evolution through the soft state. Once the accretion rate and the (X-ray) power-law distributions are homogenised for both samples, these authors find a similar scatter in the disc and corona luminosity distributions, suggesting that the (mass-scaled) common BH accretion scheme (e.g. \citealt{Merloni2003,Falcke2004,Koerding2006b}) might also hold during the soft state.

\subsection{Are accretion states identified in the LED?}\label{discuss_led}
At first glance, the distribution of objects in the LED (Figs.\,\ref{fig_led} and \ref{fig_dens}) resembles the characteristic q-shape defined by individual BHXBs when they complete a full hysteresis cycle in the HLD. This is mainly due to the narrow stripe occupied by LINER nuclei on the low-excitation side of the LED ($\rm{LyH_{IR}} \sim 0.6$--$1$), reaching sub-Eddington luminosities down to $10^{-5.6}\, \rm{L_{Edd}}$. However, the correspondence between the AGN types in Fig.\,\ref{fig_dens} and the different accretion states in the HLD is not straightforward. 

\subsubsection{The hard state} \label{hard-state}
The q-shaped object distribution found in the LED is due to the lack of high-excitation nuclei below $\sim 10^{-3}\, \rm{L_{Edd}}$, i.e. the source of high-energy photons vanishes below this limit and thus AGN lose most of their ionisation abilities (see lower panel in Fig.\,\ref{fig_sketch1}). We note that some of these nuclei are still bright in absolute terms, e.g. M87 ($L_{\rm bol} \sim 10^{42}\, \rm{erg\,s^{-1}}$; \citealt{2016MNRAS.457.3801P}), but they are low luminosity sources (likely radiatively inefficient) if we consider their large BH masses. Although the vast majority of AGN in this region of the LED belong to the LINER class, a few Seyfert nuclei are found among this low-luminosity population (e.g. NGC\,6166, NGC\,5033, NGC\,1275, NGC 5866). These Seyferts are relatively faint in [\ion{O}{iv}]$_{25.9}$ and consequently lie in the low-excitation side of the LED. From this point of view they are indistinguishable from the low-luminosity LINERs regardless of their optical classification. It is important to bear in mind that this classification is based on the classical BPT diagram \citep{1981PASP...93....5B}, which separates Seyferts and LINERs according to the emisivity of the O$^{2+}$ ($35.1\, \rm{eV}$) transition at $5007$\,\AA. However, low-luminosity Seyferts do not produce significantly higher levels of exciting radiation than faint LINERs if the O$^{3+}$ ($54.9\, \rm{eV}$) transition at $25.9\, \rm{\micron}$ is used in the diagnostics. The lack of high ionisation photons in these nuclei is consistent with the absence of the blue bump contribution to the extreme UV continuum seen in low-luminosity AGN (e.g. fig.\,7 in \citealt{Ho2008}). Instead, the range of $\rm{LyH_{IR}}$ values observed in low-luminosity AGN is in agreement with the predictions from photo-ionisation simulations assuming a power-law continuum ($F_\nu \propto \nu^{-0.4}$; light- and dark-green areas in Fig.\,\ref{fig_led}) similar to that observed in the case of NGC\,1052, where no sign of thermal-like emission is detected up to $\sim 2600$\,\AA \ (blue solid line in Fig.\,\ref{fig_input}).

Another aspect to be considered is the possible contamination of LyH$_{IR}$ due to the star formation contribution to the [\ion{Ne}{ii}]$_{12.8}$ emission \citep{2007ApJ...658..314H}. This would shift the nuclei to the low-excitation side of the LED, making them to look (artificially) hard state sources. As a matter of fact, the [\ion{O}{iv}]$_{25.9}$/[\ion{Ne}{ii}]$_{12.8}$ ratio is usually adopted as a measure of the relative AGN/star formation contribution \citep{2002A&A...393..821S}. To test this scenario we derived star formation rates for the galaxies in our sample from the flux of the polycyclic aromatic hydrocarbon (PAH) band at $11.3\, \rm{\micron}$ \citep{2019ApJ...884..136X}. PAH bands are not sensitive to AGN heating because they are dissociated in the proximity of active nuclei, and thus are typically used as a tracer of the star formation component for AGN host galaxies \citep{2012ApJ...746..168D}. PAH fluxes were compiled from \textit{Spitzer} observations in the literature (see Table\,\ref{tab_sample}), and are extracted from an aperture size equivalent to that of the fine-structure line fluxes. Fig.\,\ref{fig_sfr} shows that the impact of star formation on low-luminosity AGN is negligible due to the very low star formation rates measured in these galaxies, which have a median rate of $0.09\, \rm{M_\odot \, yr^{-1}}$ for nuclei below $10^{-3}\, \rm{L_{Edd}}$. This is not surprising, since LINERs are typically found in elliptical and early-type galaxies \citep{1997ApJ...487..568H}. Thus, the high $\rm{LyH_{IR}}$ values measured in low-luminosity AGN are not driven by star formation activity.

\begin{figure}
	\includegraphics[width=\columnwidth]{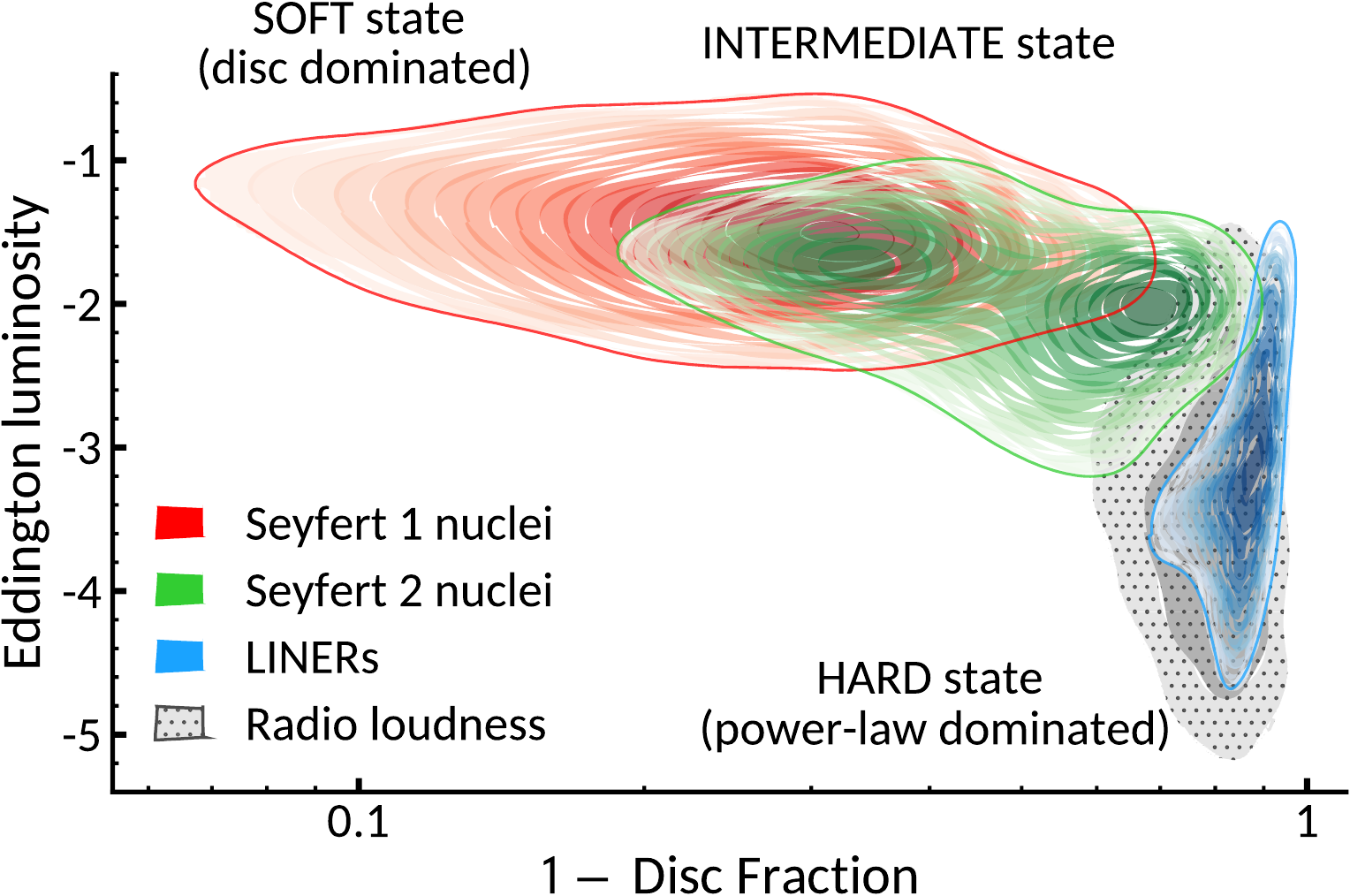}
    \caption{The main accretion states identified in the LED are compared with the density distributions, above the 50\% percentile, of the different AGN types: Seyfert 1 (red), Seyfert 2 (green), and LINER (blue). For the sake of clarity the Sy1h contours are not included here, but instead can be found in Fig.\,\ref{fig_dens}. The density distribution of the whole sample weighted by the radio loudness parameter, $\log(F_{\rm rad}/F_{\rm X})$, is shown as a grey dotted area ($> 50$\,th percentile in light grey; $> 90$\,th percentile in dark grey).}\label{fig_sketch2}
\end{figure}

Following the parallelism between X-ray binaries and supermassive BHs, a further key element is the coupling of the jet activity with the accretion state. The former are bright radio emitters during the hard state as the jet becomes the dominant energy output channel. This seems to be also the case in supermassive BHs, given that low-luminosity AGN are by far the radio-loudest sources in our sample, some of them being well-known and unambiguous examples of jet-dominated nuclei (e.g. M87 in \citealt{1999AJ....117.2185P,2016MNRAS.457.3801P}; NGC\,1052 in \citealt{2019MNRAS.485.5377F}). Although it is commonly acknowledged that LINERs are radio-loud objects \citep[e.g.][]{2003ApJ...583..145T}, the LED in Fig.\,\ref{fig_led} reveals that low-luminosity Seyferts share, along with the low excitation, the radio-loudness of these nuclei. In Fig.\,\ref{fig_sketch2} we show the probability density distributions for the different AGN types (from Fig.\,\ref{fig_dens}) compared to the density distribution of the AGN sample as a whole, weighted by the radio loudness parameter (grey dotted area). Most of the radio-loud activity in AGN seems to occur at low Eddington luminosities, with some additional contribution from low-excitation, high-luminosity nuclei in the intermediate state (Section\,\ref{intermediate}). This is in agreement with low-luminosity AGN following the fundamental plane of BH accretion \citep{Merloni2003,Falcke2004,Koerding2006b,gueltekin2009}, one of the most distinctive characteristics of accreting BHs in the hard state.

Considering all the above, we identify low-excitation and low-luminosity AGN as accreting BHs in the hard state. This group mostly comprises faint LINERs (in agreement with \citealt{Koerding2006}), but essentially the same accretion properties are shared by a few Seyferts of both types according to the LED.

\subsubsection{The soft state}
Sy1 and Sy1h nuclei are mainly found in the high luminosity ($\sim 0.001$--$1\, \rm{L_{Edd}}$) and high excitation ($\rm{LyH_{IR}} \lesssim 0.5$) part of the LED, in agreement with the presence of a prominent blue bump in their ionising continuum. This is supported by the photo-ionisation simulations including a standard accretion disc (purple shaded area in Fig.\,\ref{fig_led}), which reproduce the low $\rm{LyH_{IR}}$ values measured in these objects. Moreover, luminous and highly excited nuclei tend to be more radio quiet, as shown by the coloured symbols in Fig.\,\ref{fig_led}, in agreement with  \citet{Koerding2006} and \citet{2017A&A...603A.127S}. These characteristics suggest that AGN in the soft state are found within the loci defined by Sy1 and Sy1h galaxies in Fig.\,\ref{fig_dens}, when the disc emission dominates the radiative output and the relative jet contribution is much lower.

On the other hand, Sy2 nuclei present a bimodal distribution along the excitation axis in the LED (Fig.\,\ref{fig_dens}). Despite the lack of broad lines in their optical spectra, about half of the Sy2 nuclei behave like type 1 AGN regarding their luminosity and excitation. This suggests that the Lyman continuum and accretion luminosity of these Sy2s is as powerful as that of unobscured Seyferts, but they miss a polarised or IR detection of broad lines that has been otherwise seen in Sy1hs. Obscuration by dust in the nucleus and/or in the host galaxy could explain the type 1/type 2 dichotomy for these nuclei, which are indistinguishable in the LED. Thus, we consider Sy1s, Sy1hs, and high-excitation Sy2s as soft state AGN.

\subsubsection{The bright hard and intermediate states}\label{intermediate}
A third group of sources can be identified at low excitation and high luminosities in the LED, mainly formed by the remaining half of Sy2 nuclei (see above) and bright LINERs. However, a few Sy1s and Sy1hs are also present. Sy2 nuclei in the low-excitation peak present distinctive characteristics that cannot be merely explained by obscuration effects. Instead, the behaviour of low-excitation Sy2s in the LED mimic that of bright LINERs. The most remarkable aspect shared by most AGN in this part of the diagram is arguably the high star-formation rates found in the nuclear region of their host galaxies (e.g. \citealt{2003MNRAS.346.1055K} and \citealt{2004MNRAS.355..273C} for Sy2s, \citealt{2006ApJ...653L..13S} for LINERs), as shown by the probability density contours in Fig.\,\ref{fig_sfr}. The latter are forming stars at a median rate of $\sim 6\, \rm{M_\odot\,yr^{-1}}$ with some extreme examples at $> 100\, \rm{M_\odot\,yr^{-1}}$, such as the interacting galaxies NGC\,6240 \citep{2003A&A...409..867L} and UGC\,5101 (see Fig.\,\ref{fig_led}; \citealt{2000AJ....119..991S}). These values contrast sharply with the $\lesssim 0.09\, \rm{M_\odot\,yr^{-1}}$ found among the host galaxies of low-luminosity AGN, mostly ellipticals \citep[e.g.][]{2003MNRAS.346.1055K}, while high-excitation Seyferts (LyH$_{\rm IR} \lesssim 0.5$) show intermediate values of $\sim 2\, \rm{M_\odot\,yr^{-1}}$. Although Sy2 galaxies are known for having, on average, higher star formation rates than Sy1s \citep{1995ApJ...446..561M,2006AJ....132..401B}, the LED in Fig.\,\ref{fig_sfr} suggests that this characteristic may only be ascribed to low-excitation Sy2 nuclei, while high-excitation Sy2s behave like Sy1s (in line with the results of \citealt{2010ApJ...709.1257T}).

In galaxies with intense bursts, the LyH$_{\rm IR}$ values could be affected by the star formation contribution to the [\ion{Ne}{ii}]$_{12.8}$ line. For instance, a starburst of $10\, \rm{M_\odot\,yr^{-1}}$ could account for $90\%$ of the [\ion{Ne}{ii}]$_{12.8}$ emission (using the [\ion{Ne}{ii}]$_{12.8}$--star formation rate relation by \citealt{2019ApJ...873..103Z}), shifting the LyH$_{\rm IR}$ values from $0.25$ (i.e. $\sim$ Sy1 locus) to $0.7$ (i.e. $\sim$ low-excitation Sy2s). However, a starburst of $1\, \rm{M_\odot\,yr^{-1}}$ would have a negligible impact (from LyH$_{\rm IR} \sim 0.6$ to $0.7$). We note that only seven out of the 39 Sy2s have star formation rates above $10\, \rm{M_\odot\,yr^{-1}}$, and therefore the overall distribution of type 2 nuclei cannot be simply explained by contamination of the [\ion{Ne}{ii}]$_{12.8}$ line. This would not either explain why Sy1 and Sy1h nuclei are significantly more scarce than Sy2s in the low-excitation part of the LED, since (a priori) their spiral host galaxies could be similarly affected by bursts of star formation. This result suggests that a significant fraction of the Sy2 population present intrinsic differences with the type 1 sources that cannot be explained in terms of orientation-dependent obscuration, in agreement with \citet{2006ApJ...639..740B}, and \citet{2020ApJ...896..108Z}.

Overall, low-excitation Seyferts and bright LINERs are less radio loud than low-luminosity AGN, but they are still louder than high-excitation Seyferts. This is shown by the $50$\,th percentile contour of the radio-loudness density distribution (light grey area in Fig.\,\ref{fig_sketch2}), which includes most of the LINER population and extends almost to the low-excitation peak of the Sy2 distribution. These characteristics appear to be consistent with those of BHXBs in the bright-hard and intermediate states. Sources in the bright-hard state have increased their luminosity moving up along vertical tracks in the HLD and show luminous radio cores from a compact jet. The latter gradually fades during the intermediate state, while the accretion disc becomes increasingly dominant \citep[e.g.][]{Fender2016}, in agreement with the location of the bright LINERs and the low-excitation Seyferts in the LED. This result is also consistent with the weaker coronal line emission detected in Sy2 nuclei when compared to Sy1s of a similar bolometric luminosity \citep{2017MNRAS.467..540L}, given that coronal lines are very sensitive to the hardest part of the ionising continuum ($\gtrsim 100\, \rm{eV}$). In BHXBs, the intermediate state comprises the horizontal evolution at roughly constant luminosity with increasing disc contribution towards the soft state and is characterised by discrete and powerful relativistic radio ejections (e.g. \citealt{Bright2020} for a recent example).

\subsubsection{Radio galaxies in the LED}
The majority of radio-loud Seyfert nuclei represented by dark-filled symbols in Fig.\,\ref{fig_led} correspond to bright high-excitation radio galaxies, such as 3C\,120, 3C\,273, and Cygnus\,A. These objects typically show enormous radio lobe structures extending several to hundreds of kiloparsecs beyond the boundaries of the host galaxy \citep{1974MNRAS.167P..31F}. The cores of these galaxies are, however, not as loud as those of low-luminosity AGN when their radio continuum and X-ray emission are compared. We note that in some of these nuclei the [\ion{O}{iv}]$_{25.9}$ emission might be boosted due to the jet-ISM interaction, as seems to be the case of 3C\,120 \citep{2004ApJ...615..156S,2017ApJ...838...16T}. However, the current accretion state in the core of these galaxies might be completely decoupled from the past activity that created the enormous radio lobe structures. The dichotomy between low- and high-excitation radio galaxies has been extensively discussed in the literature \citep[e.g.][]{2010A&A...509A...6B,2012MNRAS.421.1569B} and has been associated with different disc contributions and accretion rates \citep{2019ApJ...870...53B}.

\subsection{Are state transitions expected in AGN?}\label{discuss_transit}
The identification of accretion states with similar properties for both BHXBs and supermassive BHs (e.g. the radio loud/quiet dichotomy associated with the hard and soft states) indicates the existence of a common, scale-invariant framework to describe the accretion process onto BHs across the mass range. However, this does not necessarily implies that supermassive BHs are bound to follow the standard evolution widely seen in BHXB outbursts, involving (fast) hard-to-soft and soft-to-hard state transitions. We note that the physical conditions in supermassive BHs may depart significantly from those present in BHXBs, where the (gas pressure dominated) standard accretion disc solution \citep{Shakura1973} provides a remarkably good description of the observations \citep[e.g.][]{Done2007}. For instance, the structure of the cooler and lower density AGN discs might be supported by radiation and magnetic pressure rather than gas pressure (\mbox{\citealt{1999PASP..111....1K}}, \mbox{\citealt{2019ApJ...885..144J}}).


Nevertheless, there is growing observational evidence supporting the existence of state changes in supermassive BHs. The changing look AGN phenomena proves that supermassive BHs are indeed able to develop drastic transitions, including extreme variability and the appearance or disappearance of the blue featureless continuum and the broad emission lines \citep{2016MNRAS.457..389M}. For instance, an on/off switch of the nuclear continuum source was proposed by \citet{1984MNRAS.211P..33P} to explain the spectral transitions in NGC\,4151, a Seyfert nucleus that has been changing its spectral type back and forth over the past $\sim 80$ years \citep{1943ApJ....97...28S,2008A&A...486...99S}. These transformations affect also to the hardest part of the ionising continuum, as shown by the appearance of coronal lines and a soft X-ray excess in a sample of LINER nuclei that recently became quasars \citep{2019ApJ...874...44Y,2019ApJ...883...31F}. Furthermore, changing look AGN exhibit an inflection point at $\sim 3 \times 10^{-3}$--$0.01$\,\ledd \ in the evolution of the continuum spectral slope as a function of the luminosity, drifting between softening and hardening trends \citep{2018MNRAS.480.3898N,2019ApJ...883...76R,2019arXiv190904676R}, similar to the case of X-ray binaries \citep[e.g.][]{ArmasPadilla2011,Plotkin2013,Wijnands2015}. The accretion disc-jet connection has also been discussed in the context of the appearance of radio knots following (X-ray) flux drops in bright radio galaxies, such as 3C\,120 \citep{Marscher2002,Marscher2005}. Further evidences supporting the existence of accretion state transitions in supermassive BHs have been recently found in tidal disruption events \citep{2021arXiv210104692W}. On a greater picture, AGN are though to represent a transient phase in the evolution of every massive galaxy such as the Milky Way \citep[e.g.][]{Silk1998,2000ApJ...529..745F}. Therefore, an overall active to quiescence transition is somehow expected. This should occur at some point in between the peak of AGN accretion rate density (at $z \sim 1$--$2$) and the present time \citep{Ueda2003,Shankar2009,Delvecchio2014,Aird2015}. In addition to the hysteresis pattern, one of the most universal features observed in BHXBs is the luminosity of the soft-to-hard transition ($\sim 0.01\, \rm{L_{Edd}}$; e.g. \citealt{Maccarone2003}), when the disc becomes progressively cooler, and the hot corona and the radio jet are newly observed. Contrastingly, the hard-to-soft transition might occur at different luminosities even for different outbursts of the same system \citep[e.g.][]{Dunn2010}. This is roughly consistent with the AGN picture shown by the LED, although the hardening might occur at $\sim 10^{-3}$\,\ledd, which coincides with the disappearance of the blue bump \citep{Ho2008} and the value at which the broad-line region is expected to vanish \citep{2003ApJ...589L..13N,2009ApJ...701L..91E}. The lack of high-excitation nuclei below $10^{-3}$\,\ledd \ and the similar accretion/ejection coupling support to some extent the existence of a soft-to-hard transition (or evolution) in supermassive BHs.

\subsubsection{Accretion states in the context of the current galaxy evolution paradigm}
Finally, we generally explore how the presence of different accretion states fits into the current galaxy evolution picture. The fact that most LINERs are found in passive galaxies and most Seyferts are in star-forming spirals could be understood as an argument supporting the coupling between the accretion state of the AGN and the evolution of the galaxy. This tentative evolutionary sequence would start with strong star-forming galaxies that host low-excitation Seyferts and bright LINERs located in the upper-right side of the LED (i.e. bright hard and intermediate state; Fig.\,\ref{fig_sfr}). High-excitation Seyferts (i.e. ``soft state'' AGN) are preferentially found in the \textit{green valley} between active star-forming galaxies (blue cloud) and passive early-type galaxies (red and dead sequence; \mbox{\citealt{2014MNRAS.440..889S}}, \mbox{\citealt{Heckman2014}}). The latter are at the end of their evolution, once most of the available gas has been transformed into stars. Interestingly, these are the preferred hosts of low-luminosity (hard state) AGN \citep{1996ApJ...462..183H}. This picture is also consistent with the apparent lack of chemical evolution in the broad- and narrow-line regions of quasars up to $z \sim 7$ \citep{nagao2006,juarez2009,onoue2020}. In the evolutionary sequence proposed above, the ``soft state'' quasar phase arrives after a bright ``hard state'' phase with intense star formation activity in the galaxy, which would become chemically enriched before the start of the quasar phase.

Nevertheless, this picture does not accommodate the role of changing look AGN in the evolution of their host galaxies. Besides the overall correlation found between AGN states and star formation activity described in this work, changing look transitions occur on much shorter timescales, from days to years, while star formation in galaxies is typically quenched in $\lesssim 1\, \rm{Gyr}$ \citep{2014MNRAS.440..889S}. Furthermore, some of these nuclei return to their original state in less than a few tens of years after the first transition (e.g. NGC\,4151 in \citealt{2008A&A...486...99S}; Mrk\,1018 in \citealt{2018ApJ...861...51K}; Mrk\,590 in \citealt{2018ApJ...866..123M}; NGC\,1566 in \citealt{2019MNRAS.483..558O} and \citealt{2019MNRAS.483L..88P}; NGC\,1365 in \citealt{2005ApJ...623L..93R}), suggesting that this phenomenon has likely an episodic nature with numerous back and forth changes. Thus, it is not straightforward to understand the impact of each event on their host galaxies, beyond the innermost few parsecs. This would require a comprehensive characterisation of the changing look phenomenon in order to determine key observables, such as the characteristic frequency and amplitude of these changes during the AGN duty cycle.

\vspace{0.2cm}




\section{Summary}\label{sum}
We have attempted to identify accretion states, similar to those observed in X-ray binaries, in a sample of 167 AGN. This database covers a wide range in luminosity and includes several members of each of the (optically defined) main AGN types. To this end, we developed a luminosity-excitation diagram (LED) in analogy with the hardness-luminosity diagram applied to BHXBs. Using the high- and low-excitation mid-IR fine-structure lines of [\ion{O}{iv}]$_{25.9}$ and [\ion{Ne}{ii}]$_{12.8}$, respectively, the LED shows the total line luminosity (normalised to \ledd) as a function of the so called Lyman hardness, which probes the slope of the primary ionising continuum and therefore the dominance of the big blue bump, the signature of the accretion disc. The main conclusions of this study are:

\begin{itemize}
    \item The AGN sample outlines in the LED the characteristic q-shaped morphology observed in the hardness-luminosity diagram of individual X-ray binaries during a typical hysteresis cycle.
    \item Below $\lesssim 10^{-3}\, \rm{L_{Edd}}$, AGN are found in the hard state. This is indicated by the low excitation gas traced by the [\ion{O}{iv}]$_{25.9}$ and [\ion{Ne}{ii}]$_{12.8}$ lines, consistent with a negligible contribution from the accretion disc to the ionisation continuum, and confirmed by the radio loudness characteristic of low-luminosity AGN.
    \item Seyferts with broad lines in their spectra (Sy1s and Sy1hs) are identified as soft state nuclei, in agreement with previous studies. These have gas excitation consistent with being dominated by the blue bump (i.e. accretion disc) emission, and they are mostly radio quiet.
    \item Sy2 nuclei show a bimodal distribution along the excitation axis. The high-excitation group is indistinguishable from Sy1 and Sy1h nuclei in the LED, showing similar excitation and accretion luminosities.
    \item Low-excitation Sy2 nuclei and bright LINERs show similar properties and can be tentatively associated with the bright hard and intermediate states. The host galaxies of these nuclei are characterised by very high star formation rates in their nuclear regions ($\sim 6\, \rm{M_\odot\,yr^{-1}}$).
    \item The traditional AGN classification based on the optical emission line properties (Sy1/Sy2/LINER) does not provide an accurate description of the accretion state in supermassive BHs. This seems to be case of at least the Sy2 and LINER types. In particular, the latter class is known to be very heterogeneous, covering a very wide range in luminosity, star formation rate, and host galaxy morphology.
\end{itemize}

Finally, we remark that IR spectroscopic observatories, such as the \textit{James Webb Space Telescope} \citep{2006SSRv..123..485G} in the short term, and a future generation of cryogenic IR observatories in the long term (e.g. \textit{SPICA}, \citealt{2018PASA...35...30R}; the \href{https://asd.gsfc.nasa.gov/firs/docs/OriginsAPCwhitepaperassubmitted10July2019.pdf}{\it Origins Space Telescope}\footnote{\url{https://asd.gsfc.nasa.gov/firs}}), should play a crucial role in the study of accretion states in supermassive BHs, since they will offer a unique probe of the primary ionising AGN continuum, which is otherwise inaccessible by other means.

\section*{Acknowledgements}
The authors would like to thank Dr. M. Armas Padilla for her helpful suggestions and Dr. L. Spinoglio for his thoughtful revision and discussion of the draft. The authors would also like to thank the referee for carefully reading the manuscript and providing constructive comments that helped to improve the quality of the paper. JAFO acknowledges financial support by the Agenzia Spaziale Italiana (ASI) under the research contract 2018-31-HH.0. TMD acknowledges support from the State Research Agency (AEI) of the Spanish Ministry of Science, Innovation and Universities (MCIU) and the European Regional Development Fund (ERDF) under grant AYA2017-83216-P. TMD acknowledges support from the Consejer\'ia de Econom\'ia, Conocimiento y Empleo del Gobierno de Canarias and the ERDF under grant ProID2020 010104. TMD acknowledges support from the Ram\'on y Cajal Fellowship RYC-2015-18148. We acknowledge the usage of the HyperLeda database\footnote{\url{http://leda.univ-lyon1.fr}} and Astropy\footnote{\url{http://www.astropy.org}}, a community-developed core Python package for Astronomy \citep{astropy2013,astropy2018}. This research has also made use of the NASA/IPAC Extragalactic Database (NED), which is operated by the Jet Propulsion Laboratory, California Institute of Technology, under contract with the National Aeronautics and Space Administration.


\section*{Data Availability}

The data underlying this article are available in the article and in its online supplementary material.



\bibliographystyle{mnras}
\bibliography{agn_led,Libreria}





\bsp	
\label{lastpage}
\end{document}